\documentclass[]{aa}

\usepackage{tikz}
\usepackage{xcolor}

\usepackage{pgf}	

\usepackage{txfonts}
\usepackage{natbib}
\usepackage{epsf}
\usepackage{rotating}
\usepackage{graphics}
\usepackage{verbatim}
\usepackage{longtable}

\newcommand{\logR}{$\log{(R'_{\rm HK})}$}

\def \MJ{M$_{\mathrm{Jup}}$}
\def \ME{M$_\oplus$}

\def \msol{M$\mathrm{_\odot}$}
\def \kms{km\,s$^{-1}$}
\def \ms{m\,s$^{-1}$}
\def \1s{$1\,\sigma$}

\def \t0{T$_0$}

\def \sini{\sin i_\star}

\newcommand{\teta}{\boldsymbol{\theta}}
\newcommand{\evid}{\mathcal{Z}}
\newcommand{\nodata}{$--$}

\newcommand{\vc}{v_c(t_i)} 
\newcommand{\vp}{v_p(t_i;\,\pi, \mu, \delta)}
\newcommand{\vkep}{v_k(t_i; \mathbf{\theta})}

\newcommand{\struct}{\epsilon}

\newcommand{\emcee}{{\tt emcee}}

\DeclareUnicodeCharacter{2212}{ --}

\begin{document}

\title{The SOPHIE search for northern extrasolar planets\thanks{Based on observations collected with the {\it SOPHIE}  spectrograph on the 1.93-m telescope at Observatoire de Haute-Provence (CNRS), France by the {\it SOPHIE}  Consortium.}\thanks{Table \ref{table.rv} is only available in electronic form
at the CDS via anonymous ftp to cdsarc.u-strasbg.fr (130.79.128.5)
or via http://cdsweb.u-strasbg.fr/cgi-bin/qcat?J/A+A/}}

\subtitle{XIV. A temperate ($T_\mathrm{eq}\sim 300$ K) super-earth around the nearby star Gliese 411.}

\author{
R.~F.~D\'iaz\inst{1,2} \and
X.~Delfosse\inst{3}\and
M.~J.~Hobson\inst{4}\and
I.~Boisse\inst{4} \and
N.~Astudillo-Defru\inst{5}\and
X.~Bonfils\inst{3}\and 
G.~W.~Henry\inst{6}\and
L.~Arnold\inst{7}\and
F.~Bouchy\inst{8}\and
V.~Bourrier\inst{8}\and
B.~Brugger\inst{4}\and
S.~Dalal\inst{9}\and
M.~Deleuil\inst{4}\and
O.~Demangeon\inst{10, 11}\and
F.~Dolon\inst{7}\and
X.~Dumusque\inst{8}\and
T.~Forveille\inst{3}\and 
N.~Hara\inst{8}\and
G.~H\'ebrard\inst{9,7} \and 
F.~Kiefer\inst{9}\and
T.~Lopez\inst{4}\and
L.~Mignon\inst{3}\and
F.~Moreau\inst{7}\and
O.~Mousis\inst{4}\and
C.~Moutou\inst{4,12}\and 
F.~Pepe\inst{8}\and
S.~Perruchot\inst{7}\and
Y.~Richaud\inst{7}\and
A.~Santerne\inst{4} \and
N.~C.~Santos\inst{10, 11}\and
R.~Sottile\inst{7}\and
M.~Stalport\inst{8}\and
D.~S\'egransan\inst{8}\and 
S.~Udry\inst{8}\and
N.~Unger\inst{1,2}\and
P.~A.~Wilson\inst{13,14}
}

 \offprints{R.F D\'iaz (rodrigo@iafe.uba.ar)}

\institute{
Universidad de Buenos Aires, Facultad de Ciencias Exactas y Naturales. Buenos Aires, Argentina.\and
CONICET - Universidad de Buenos Aires. Instituto de Astronom\'ia y F\'isica del Espacio (IAFE). Buenos Aires, Argentina.\and
Univ. Grenoble Alpes, CNRS, IPAG, 38000 Grenoble, France.\and
Aix Marseille Univ, CNRS, CNES, LAM, Marseille, France.\and
Departamento de Astronom\'ia, Universidad de Concepci\'on, Casilla 160-C, Concepci\'on, Chile.\and
Center of Excellence in Information Systems, Tennessee State University, Nashville, TN 37209, USA.\and
    Observatoire de Haute Provence, CNRS, Aix Marseille Universit\'e, Institut Pyth\'eas UMS 3470, 04870 Saint-Michel-l'Observatoire, France.\and 
Observatoire Astronomique de l'Universit\'e de Gen\`eve,  51 Chemin des Maillettes, 1290 Versoix, Switzerland.\and
Institut d'Astrophysique de Paris, UMR7095 CNRS, Universit\'e Pierre \& Marie Curie, 98bis boulevard Arago, 75014 Paris, France.\and 
Instituto de Astrof\'isica e Ci\^encias do Espa\c{c}o, Universidade do Porto, CAUP, Rua das Estrelas, 4150-762 Porto, Portugal.\and
Departamento de F\'isica e Astronomia, Faculdade de Ci\^encias, Universidade do Porto, Rua do Campo Alegre, 4169-007 Porto, Portugal. \and
Canada-France-Hawaii Telescope Corporation, 65-1238 Mamalahoa Hwy, Kamuela, HI 96743, USA.\and
Department of Physics, University of Warwick, Coventry CV4 7AL, UK.\and
Centre for Exoplanets and Habitability, University of Warwick, Coventry CV4 7AL, UK.
}

\date{Received TBC; accepted TBC}
      
\abstract{Periodic radial velocity variations in the nearby M-dwarf star Gl411 are reported, based on measurements with the SOPHIE spectrograph. Current data do not allow us to distinguish between a 12.95-day period and its one-day alias at 1.08 days, but favour the former slightly. The velocity variation has an amplitude of 1.6 \ms, making this the lowest-amplitude signal detected with SOPHIE up to now. We have performed a detailed analysis of the significance of the signal and its origin, including extensive simulations with both uncorrelated and correlated noise, representing the signal induced by stellar activity. The signal is significantly detected, and the results from all tests point to its planetary origin. Additionally, the presence of an additional acceleration in the velocity time series is suggested by the current data. On the other hand, a previously reported signal with a period of 9.9 days, detected in HIRES velocities of this star, is not recovered in the SOPHIE data. An independent analysis of the HIRES dataset also fails to unveil the 9.9-day signal.

If the 12.95-day period is the real one, the amplitude of the signal detected with SOPHIE implies the presence of a planet, called Gl411 b, with a minimum mass of around three Earth masses, orbiting its star at a distance of 0.079 AU. The planet receives about 3.5 times the insolation received by Earth, which implies an equilibrium temperature between 256 K and 350 K, and makes it too hot to be in the habitable zone. At a distance of only 2.5 pc, Gl411 b, is the third closest low-mass planet detected to date. Its proximity to Earth will permit probing its atmosphere with a combination of high-contrast imaging and high-dispersion spectroscopy in the next decade.}

\authorrunning{D\'iaz et al.}
\titlerunning{A super-Earth around the nearby star Gl411}

\keywords{planetary systems -- techniques: radial velocities -- stars: low-mass -- stars: individual: \object{Gl411}}
\maketitle

\section{Introduction \label{sect.intro}}

It is usually stated that low-mass stars are prime targets for extrasolar planet searches. Their relative low masses and small radii permit detecting smaller planets than on their hotter counterparts. M dwarfs also present the advantage that planets orbiting in the habitable zone are much easier to detect, because in addition to the advantages due to their masses and radii, the habitable zone lies much closer to the star than in solar-type stars. In addition, the atmospheres of planets transiting close M dwarfs could be amenable to characterisation in the coming decade by the James Webb Space Telescope \citep{doyon2014, beichman2014}, or on the ground with stabilised high-resolution spectrographs on medium to large-size telescopes \citep[e.g.][]{snellen2008, wyttenbach2017, allart2018}. Another favourable aspect is that the majority of the closest neighbours to the solar system are low-mass stars. This opens the possibility of the characterisation of the atmospheres of non-transiting planets around low-mass stars by a combination of high-dispersion spectroscopy and high contrast imaging \citep{sparksford2002, snellen2015}.  These observations will be plausible with the next-generation large-aperture telescopes, such as the E-ELT, but even with the current instrumentation, this is possible for the most favourable cases \citep{lovis2017}. Similar planets around solar type stars will only be accessible for characterisation with highly ambitious space missions, which will not be in operation before the middle of the century.

Besides theoretical advantages, recent results show that rocky planets orbiting low-mass stars are abundant: based on results from the HARPS velocity survey, \citet{bonfils2013} reported an occurrence rate above 50\% for super-Earth planets on orbits with periods between 10 and 100 days; using Kepler photometry \citet{dressingcharbonneau2015} reported an average of 2.5 planets with radii between one and four Earth radii per M-star host, on orbits with periods shorter than 200 days; \citet{gaidos2016} concluded that M stars host an average of 2.2 planets per star with periods between 1.5 and 180 d. Furthermore, planets orbiting in the habitable zone of low-mass stars are also relatively common. \citet{dressingcharbonneau2015} reported an average between 0.16 - 0.24 Earth-sized planets in the habitable zone of M dwarfs, based on Kepler photometry.  \citet{bonfils2013} reported a fraction $0.41^{+0.54}_{-0.13}$, although this number is probably slightly overestimated, as it included a planet candidate later shown to be likely produced by activity \citep{robertson2014}. 

The SOPHIE search for northern extrasolar planets is a radial velocity survey operating since 2006 \citep{bouchy2009}, and is divided in a number of sub-programmes, each targeting different stars and/or planets. In particular, the sub-programme three (SP3) is dedicated to finding planets around low-mass stars. A sample of relatively bright northern M-dwarf stars is observed systematically in order to constrain the presence of planets and also to find targets amenable to detailed characterisation, either now or using future observing facilities.

Gl411 is one of the brightest M-dwarf stars in the sky \citep{lepinegaidos2011, gaidos2014}. As such, it has been intensively observed for many years. In a number of opportunities, detections of orbiting companions were reported, starting with the astrometric study by \citet{vandeKamp1951}, who reported a 1.14-yr period object on an eccentric orbit, with a minimum mass as low as 0.03 \msol. The result was later revised by \citet{lippincott1960}, who concluded that the orbiting companion is on an eigth-yr orbit, with a mass as low as 0.01 \msol, that is around ten Jupiter masses. More recently, \citet{gatewood1996} reported the detection of an astrometric signal with a period of 5.8 years, which implies a companion mass as low as 0.9 \MJ. These detections remain unconfirmed to this day, as far as we are aware. Using radial velocity observations from the HIRES spectrograph, \citet{butler2017} recently reported the presence of a periodic signal with a period of 9.9 days, and an amplitude compatible with a planetary companion with a minimum mass of 3.8 \ME.

In this article, we report the detection of a 3-Earth-minimum-mass companion on a 12.95-day period orbit around the nearby star \object{Gl411} (\object{Lalande 21185}, \object{HD 95735}), based on almost seven years of measurements from the SOPHIE SP3. 
Interestingly, Gl411 was initially chosen as a possible SOPHIE standard star because of its brightness and of its reported relative low variability level of 10 \ms \citep{butler2004}. Before the upgrade of SOPHIE in 2011 (see Sect.~\ref{sect.observations}), the low velocity scatter was confirmed and therefore Gl411 was continued to be observed as a standard star. The low velocity rms was further confirmed during our first season of observation with the upgraded instrument. The consequently large number of measurements obtained on this target finally revealed a periodic 1.6-\ms variation with a period of around 12.95 days. 

In Sect.~\ref{sect.stellar} of this article, the stellar parameters of Gl411 are presented, possible values of the rotational period of the star are discussed, and the detection of the rotational modulation using photometric ground-based observations is described. In Sect.~\ref{sect.observations} we present the spectroscopic observations as well as the measurement of the radial velocities and ancillary observables. In Sect.~\ref{sect.analysis} the physical and probability model used to describe the data is presented, and the analysis of the data is described. The detection of a periodic signal, together with the validation of its planetary origin, are also presented here. In Sect.~\ref{sect.discussion} we present a few discussions on the planetary companion detected, and the analysis of the HIRES data. In particular, in Sect.~\ref{sect.hires} we show that the 9.9-day period signal announced by \citet{butler2017} is not found in the SOPHIE data, and that an independent analysis of the HIRES velocity also fails to recover the reported signal. In Sect.~\ref{sect.conclusion} we present some brief conclusions.

\section{Stellar characterisation \label{sect.stellar}}

The star Gl411 is an early M dwarf \citep[M1.9;][]{mann2015} located in the constellation of Ursa Major. It is the sixth star from the Sun, after the three-star Alpha Centauri system, Barnard's star, and CN Leonis \citep{vanaltena1995}. With apparent visual magnitude of 7.5 \citep[J=4.2;][]{2mass}, it is one of the brightest M-dwarf stars in the sky \citep{lepinegaidos2011, gaidos2014}. The stellar parameters were determined by \citet{mann2015} and are listed in Table~\ref{table.stellarparams}. The mass was computed by these authors using the mass-luminosity relation by \citet{delfosse2000} and the radius was obtained from the $T_\mathrm{eff}$ and the bolometric flux, the latter estimated from optical and near-infrared flux-calibrated spectra, and the former from comparing the spectra to the BT-SETTL atmosphere models \citep{allard2012,allard2013}.

The X-ray flux of Gl411 was measured by the XMM-Newton satellite \citep{jansen2001} and presented in the third XMM-Newton serendipitous source catalogue \citep{rosen2016}. The value of the flux measured in 2001 ($F_x = (6.59\pm0.13)\times10^{-13}$~erg~cm$^{-2}$/s), according to the DR8 of the catalogue) implies $R_x = L_x/L_\mathrm{bol} = (6.09 \pm 0.13)\times10^{-6}$, using the bolometric flux from \citet{mann2015}, and where the uncertainties were obtained by propagating the errors in the X-ray count rate, the bolometric flux and the stellar parallax, assuming normal errors in all cases. 

Regarding the rotational velocity and period, \citet{delfosse98} reported an upper limit for the velocity, $v \sini < 2.9$ \kms. Using SOPHIE spectra, \citet{houdebine2010} measured it to be $v \sini = 0.61$ \kms, with a general uncertainty of 0.3 \kms. However, this value is at the limit of their detection sensitivity. Using a value of the stellar radius based on V-band photometry, they measured a rotational period, $P/\sini = 36.4$ d. With the radius determination from \citet{mann2015}, the rotational period is $P/\sini = 32.3$ d. These determinations based on the $v \sini$ suffer from a relative uncertainty of around 51\%, if a 10\% uncertainty on the stellar radius is assumed. 

On the other hand, the rotation-activity calibration for low-mass stars of \citet{astudillodefru2017a} indicates a rotational period $P_\mathrm{rot} = 91 \pm 18$ days, roughly twice the value reported by \citet{noyes84}, for the measured value of the activity index, \logR$=-5.51\pm0.13$. The activity-rotation calibration based on X-ray luminosity by \citet{kiragastepien2007} produces a rotational period of $72.5 \pm 8.4$ days for the XMM-Newton X-ray flux, and using the "low mass" calibration, obtained using stars with masses between 0.33 \msol and 0.39 \msol. Gl411 is close to the upper bound of the calibration stars, and therefore the "medium mass" calibration of \citet{kiragastepien2007}, using stars with masses between 0.43 \msol and 0.52 \msol, may also be considered. The medium-mass calibration produces a rotational period of $57\pm 12$ days.

Finally, \citet{noyes84} reported the rotational period to be 48 days, based on modulation detected in the S-index measured by the H-K programme \citep{mw}. This value implies a rotational velocity between $v = 0.41 - 0.46$ \kms, depending on the adopted stellar radius. Because of the large uncertainties in rotational velocity and those associated with the activity calibrations, these values of rotational period are formally in agreement with each other (differences below 3 $\sigma$).

\subsection*{Photometric observations}
The stellar rotational period is a key element in the validation of the planetary nature of periodic signals detected in RV. To measure a precise rotational period, we analysed photometric observations of Gl411 acquired contemporaneously with our SOPHIE radial-velocity measurements during the 2011 through 2018 observing seasons with the Tennessee State University (TSU) T3 0.40~m automatic photoelectric telescope (APT) at Fairborn Observatory in southern Arizona. Details on the instrument, observing strategy, and data analysis are given in the Appendix~\ref{sect.apt}.

\begin{figure}
\includegraphics[width=\columnwidth]{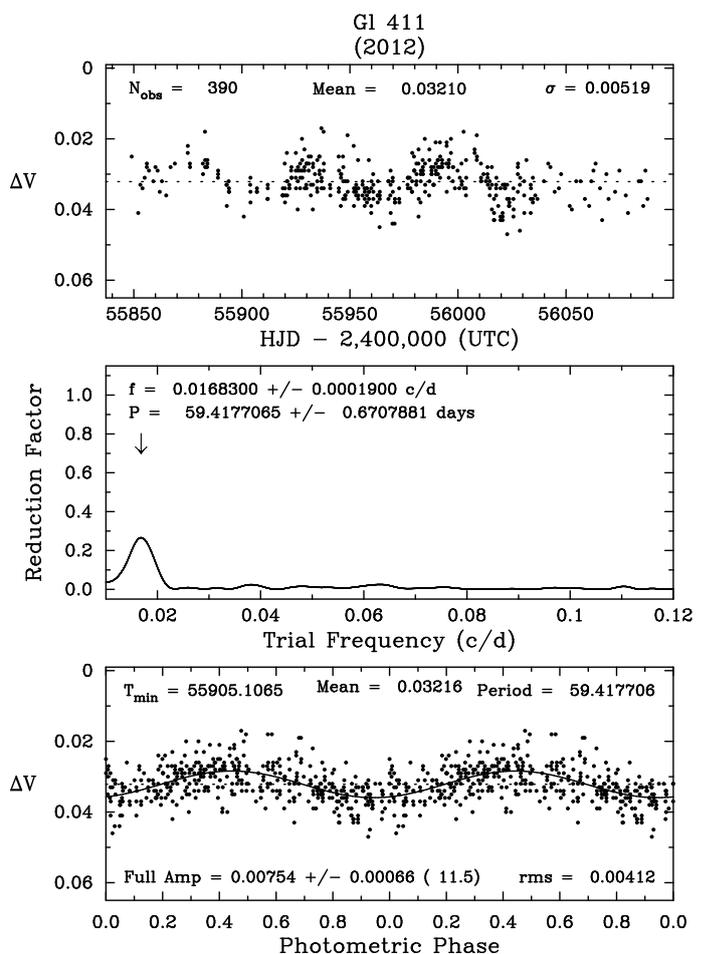}
\caption{$Top$:  Johnson $V$ photometry of Gl411 from the 2012 observing
season, acquired with the TSU T3 0.40~m APT at Fairborn Observatory.
$Middle$:  Frequency spectrum of the 2012 $V$ observations of Gl411. The 
best frequency occurs at $0.01683\pm0.00019$ cycles per day, corresponding to
a best period of $59.42\pm0.67$ days.  $Bottom$:  The 2012 $V$ observations 
phased with the best period of 59.42 days.  The phase curve shows coherent 
variability with a peak-to-peak amplitude of 0.0075~mag, which we take to 
be rotational modulation of photospheric spots.  Seven of the eight 
observing seasons exhibit similar modulation (see Table~\ref{table.photometry}).} 
\label{fig.photometry}
\end{figure}

A frequency spectrum, computed as the reduction in total variance of the data 
versus trial frequency, was performed on the $V$ and $B$ data from each of the 
eight observing seasons.  The frequency analyses spanned a frequency range of 
0.005 to 0.95 cycles/day (c/d), corresponding to a period-search range of 1 to 200 days.  The results of our analysis of the photometric data are summarized in Table~\ref{table.photometry}. The resulting periods and peak-to-peak amplitudes are given in columns 6 and 7. Periodic variations were found in seven of the eight observing seasons, at 
least for the $B$ data sets.  The weighted mean of the 12 detected periods is $56.15\pm0.27$ days.  The peak-to-peak amplitudes ranged from only 0.002 to
0.008 mag.  We take the 56.15-day period to be the stellar rotation period
revealed by rotational modulation in the visibility of small star spots on the 
stellar surface.  A sample frequency spectrum from 2012, covering a limited
part of the total frequency range, is shown in Figure~\ref{fig.photometry}.

\begin{table}
\caption{Stellar parameters of Gl 411.}
\label{table.stellarparams}
\begin{tabular}{llcc}
\hline
\hline
Parameter&&Value&Ref.\\
\hline
R.A. (J2000) && 11:03:20.19 & (1)\\

Dec. (J2000) && +35:58:11.57 & (1)\\

Proper motion \\

\hspace{2em} R.A.: $\mu_\alpha$ &[mas/yr] & $-580.27\pm0.62$ & (1)\\

\hspace{2em} Dec.: $\mu_\delta$ &[mas/yr] & $-4765.85\pm0.64$ & (1)\\


Parallax, $\pi$ &[mas] & $392.64\pm0.67$ & (1)\\

Distance, $d$ & [pc] & $2.5468 \pm 0.0043$\\


Spectral type & &M1.9 & (2)\\

Eff. temperature, $T_\mathrm{eff}$ &[K]& $3563\pm60$ & (2)\\

\logR && $-5.51\pm0.13$&(3)\\

Rot. period, $P_\mathrm{rot}$ &[d] &$56.15 \pm 0.27$&(4)\\


Mass, $M_\star$ &[$M_\odot$] & $0.386\pm0.039$ &(2)\\

Radius, $R_\star$ &[$R_\odot$] & $0.389\pm0.013$ &(2)\\

\text{[Fe/H]}&	&$-0.38\pm0.08$ &(2)\\

Luminosity, $L_\star$ &[$L_\odot$] & $0.0220\pm0.0021$\\

$L_x/L_\mathrm{bol}$ & & $(6.09 \pm 0.13)\times10^{-6}$ & (5)\\

\hline
\end{tabular}
\tablefoot{(1) \citet{vanleeuwen2007}; (2) \citet{mann2015}; (3) This work; mean value and standard deviation reported; (4) This work, based on photometric observations.; (5) based on \citet{rosen2016}}
\end{table}

\section{Observations and data reduction \label{sect.observations}}
Radial velocity measurements of Gl411 were obtained with the SOPHIE spectrograph, mounted on the 1.93-m telescope of the Observatoire de Haute-Provence, in France. SOPHIE is a fibre-fed, cross-dispersed \emph{echelle} spectrograph whose dispersive elements are kept at constant pressure. It is installed in a temperature stabilised environment \citep{perruchot2008} to provide high-precision radial velocity measurements over long timescales.

Observations were carried out in the high-resolution mode of the instrument, which provides resolving power R = 75\,000. Calibration spectra of either a ThAr lamp or a Fabry-Perot \'etalon --starting on January 2018-- were recorded simultaneously using the second optical fibre. This allows monitoring for velocity drifts in the observations with respect to the baseline wavelength calibrations, which are performed every around two hours throughout a typical observing night.

The star Gl411 was observed with SOPHIE 178 times since the beginning of the instrument operation in 2006. However, we will only use the data acquired over the 157 visits between October 2011 and June 2018, performed after the instrument upgrade in 2011, when the fibre links were partially replaced by an octagonal-section fibre \citep{perruchot2011, bouchy2013}. Velocities obtained before the fibre upgrade are affected by a systematic effect, described in detail by \citet{boisse2010, boisse2011b}, \citet{diaz2012}, and \citet{bouchy2013}, which hindered the detection of low-amplitude signals. The dispersion of the raw velocity measurements --that is without correction of the zero-point changes described below-- is 11.4 \ms and 3.9 \ms, before and after the fibre upgrade, respectively.

Exposure time was generally 900 seconds, but it was augmented in 17 visits (with a maximum of 1800 seconds) to acquire the required signal-to-noise ratio (S/N) under degraded weather conditions. The median S/N per pixel at 550 nm was 132. One visit (JD=2\,457\,728.657) had an extremely low S/N and was discarded. An additional visit (JD=2\,457\,105.475) produced a clear outlier and was also discarded, leaving a total number of 155 visits for the analysis detailed below.

Although four observations were performed with the Moon less than 30 degrees away from the target, and close to the full phase, the velocity of the Moon is always far from the measured mean velocity of the target, which is around -84.6 \kms. Therefore, no systematic effect on the SOPHIE velocities of the target expected from moonlight contamination (see Sect~\ref{sect.mcmc}).

\begin{table*}
\caption{SOPHIE measurements. Radial velocities (RV, col.~1), photon noise and calibration uncertainties ($\sigma_\mathrm{RV}$, col.~2), full width at half maximum (FWHM, col.~3), bisector velocity span (BIS, col.~4), activity index based on H$_\alpha$ line (I$_{\mathrm{H}\alpha}$, col.~5), activity index based on Ca II H \& K lines (\logR, col.~6), interpolated zero-point offset (col.~7). Full version available online.}
\label{table.rv}
\center
\begin{tabular}{llllllll}
\hline
\hline\noalign{\smallskip}
Time & RV & $\sigma_\mathrm{RV}$ & FWHM & BIS & I$_{\mathrm{H}\alpha}$ & \logR & Zero-point offset\\ 
BJD - 2\,400\,000 & [km s$^{-1}$] & [km s$^{-1}$] & [km s$^{-1}$] & [m s$^{-1}$] & -- & -- & [m s$^{-1}$]\\ 
\hline\noalign{\smallskip}
55865.677824 & -84.6092 & 0.0016 & 4.613 & 9.00 & 0.323 & -5.568 & 3.764\\ 
55875.669890 & -84.6147 & 0.0014 & 4.630 & 11.33 & 0.327 & -5.594 & 1.002\\ 
55879.706792 & -84.6149 & 0.0013 & 4.644 & 9.83 & 0.322 & -5.631 & 0.998\\ 
55906.683957 & -84.6152 & 0.0013 & 4.656 & 7.67 & 0.317 & -5.451 & 1.428\\ 
55917.706581 & -84.6132 & 0.0019 & 4.665 & 11.00 & 0.323 & -5.610 & 3.210\\ 
\hline
\end{tabular}
\end{table*}

\begin{figure*}
\center
\input{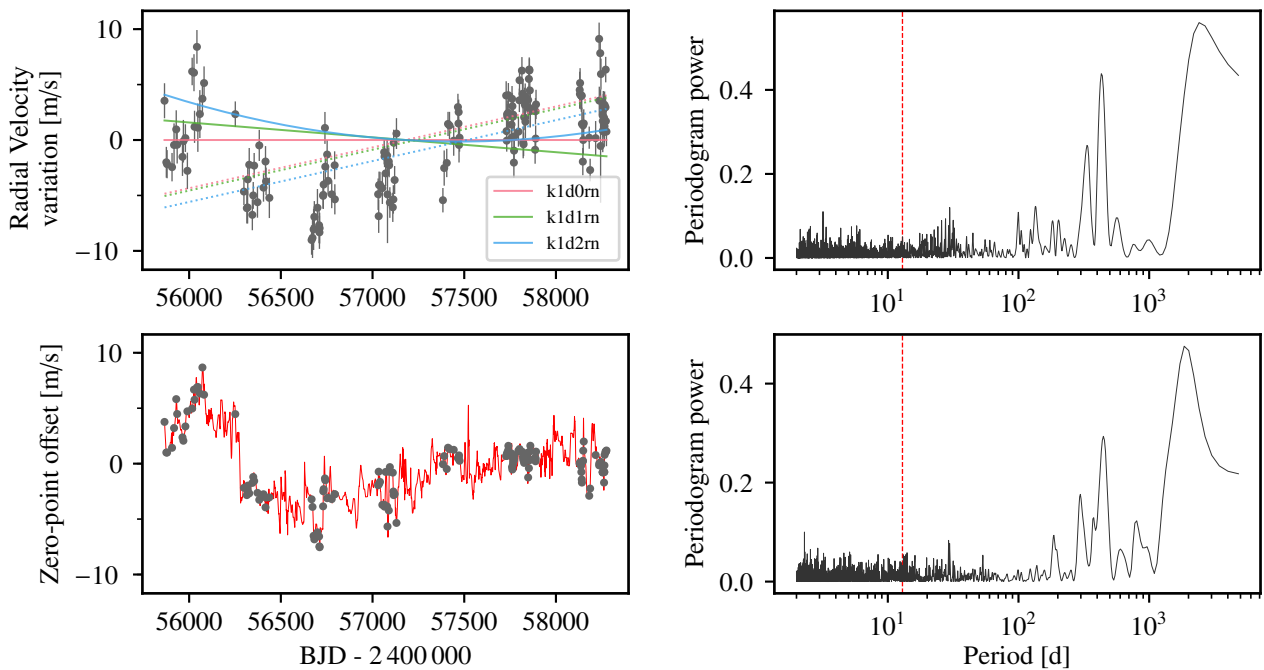}
\caption{SOPHIE velocities and zero-point offset (upper left and lower left panels, respectively) with their corresponding GLS periodograms (right panels). \emph{Upper left panel:} The solid and dashed curves represent the additional acceleration and secular (projection) acceleration terms under the three models with largest Bayesian evidence. \emph{Lower left panel:} the thin red line is the spline interpolation of the zero-point shifts measured on a number of standard stars (see text for details), while the points represent the interpolation evaluated at the epochs of the velocity measurements. \emph{Right panels:} In the periodograms of SOPHIE raw radial velocities of Gl411 and zero-point offset of the spectrograph, the vertical red lines mark the frequency of the detected periodic signal.}
\label{fig.modelintime}
\end{figure*}

The spectra were reduced using the SOPHIE pipeline described by \citet{bouchy2009}, and corrected from the charge transfer inefficiency effect using the method by \citet{bouchy2009c}. The radial velocity (RV) was measured from the extracted 2-D spectra using a template-matching procedure described in detail by \citet{astudillodefru2015, astudillodefru2017b} and by \citet{hobson2018}. The mean estimated uncertainty due to photon noise and calibration error is 1.38 \ms. The instrumental velocity drifts measured simultaneously using the second spectrograph fibre were subtracted from the measured velocities\footnote{The mean of the measured drifts is 16~cm\,s$^{-1}$ and have a standard deviation of 1.7~m\,s$^{-1}$.}. The resulting radial velocities are listed in Table~\ref{table.rv}, and plotted in the upper left panel of Fig.~\ref{fig.modelintime}.

Additional observables are obtained from the SOPHIE spectra. The extracted two-dimensional spectra are cross-correlated with a spectral mask constructed based on high-resolution HARPS spectra of Gl581, convolved with a PSF to degrade the resolution to the SOPHIE level. The full-width at half-maximum (FWHM) and bisector velocity span (BIS) \citep{queloz2001} are measured from the resulting cross-correlation function (CCF). Their uncertainties are set to twice the velocity error and to 10 \ms, for the BIS and FWHM, respectively. Besides, activity proxies based on the Ca II H \& K line fluxes and the H$\alpha$ line flux are obtained following the procedure described by \citet{astudillodefru2017a}, and \citet{boisse2009}, respectively. These data are provided as well in Table~\ref{table.rv}.

\section{Data analysis \label{sect.analysis}}

\subsection{The model \label{sect.model}}
Given an observed velocity time series, $\{t_i, v_i, \sigma_i\}$, for $i = \{1, \ldots, N\}$, the model for the velocity measurement taken at time $t_i$, $v_i$, can be written, in a very general manner, as:
$$
v_i = m_i + \epsilon_i\;\;,
$$
where $m_i$ is the model prediction at time $t_i$ and $\epsilon_i$ is the error term. The variation in the radial velocity seen in Fig.~\ref{fig.modelintime} is caused by, at least, a combination of instrumental systematics, and a velocity projection effect. These will be the first elements in our model. We will investigate the possibility that either an additional physical acceleration or the effect of an orbiting sub-stellar companion, or both, are also necessary components of the model. 

Even after the installation of the octagonal-section fibre in 2011, SOPHIE suffers from a well-observed zero-point change. This effect is monitored by a set of constant stars that are observed regularly, and make up the zero-point velocity 'master' \citep[see][for more details]{courcol2015, hobson2018}. Here, the zero-point master is constructed leaving out Gl411, which was considered up to now a constant star. Instead of subtracting the measured master from the observations directly, a linear dependence with the zero-point velocity master, $\vc$, was included in the model. In this way, variations which are covariate with the zero-point velocity changes, such as long-term velocity variations, are correctly taken into account. To be able to evaluate $\vc$ at the epochs of the observations of Gl411, we interpolate the velocities of the constant stars using quadratic splines. In the lower left corner of Fig.~\ref{fig.modelintime} the spline interpolation is plotted as a thin red line, and the values at the epochs of the observations of Gl411 are shown as grey points. In the lower right panel of Fig.~\ref{fig.modelintime}, a Generalised Lomb Scargle \citep[GLS;][]{zechmeisterkurster2009} of the interpolated zero-point values is presented. The zero-point variations are dominated by a low-frequency variability with period of around 2000 days, followed by peaks at around 450, 300, and and 190 days.

The secular acceleration produced by the changing projection of the star's velocity vector can be computed using the measured distance, proper motion and absolute radial velocity listed in Table~\ref{table.stellarparams} \citep{kurster2003}. We include this term in the model as well, $\vp$, using informative priors on the parallax and proper motion measured at the epoch of the Hipparcos catalog, J1991.25, $JD = 2\,448\,349.0625$ \citep{perryman1997}\footnote{Gl411 is not included in the latest Gaia data release, probably because of its brightness and/or its high proper motion, which are both cited as sources of incompleteness \citep[see][Sect 6.2]{gaiaDR2}.}, and fixing the star's declination, $\delta$, to the value in Table~\ref{table.stellarparams}. For the mean value of the prior distribution, the secular acceleration has an amplitude of about 1.34 \ms/yr. We will call k0d0 the model containing only these two terms, and the expression for the model prediction is then
$$
m^{k0d0}_i = a_c \vc + \vp\;\;,
$$
for model parameters $a_c$, $\pi$, and $\mu$.

We also allow for the possibility of additional variation coming from a secular acceleration, which we model by a linear (d1) or quadratic (d2) term, which produces the model:
$$
m^{k0d0}_i + \gamma_1\;(t_i - t_\mathrm{ref}) + \gamma_2\;(t_i - t_\mathrm{ref})^2\;\;,
$$
where $t_\mathrm{ref}$ is a reference time, taken as $JD = 2\,457\,180$, close to the mean time of the SOPHIE time series. The resulting model is called k0d1 if $\gamma_2=0$, and k0d2 otherwise.

Finally, we include the effect of orbiting companions. We assume the effect of multiple companions is simply the sum of the effects of the individual companions (i.e. we neglect mutual interactions between planets). We denote the effect of the individual companions at time $t_i$ as $\vkep$, where $\mathbf{\theta}$ is the model parameter vector. $\vkep$ is assumed to be a Keplerian function parametrised by the orbital period, $P$, the amplitude of the radial velocity signal, $K$, the orbital eccentricity, $e$, the argument of pericentre, $\omega$, and the mean longitude at a given epoch, $L_0$. Under this parametrisation \citep[see, e.g.][]{murraydermott2000}:
$$
\vkep = K \left[\cos\left(f_i + \omega\right) + e \cos(\omega)\right]\;\;,
$$
where $f_i = f(t_i; P, e, \omega, L_0)$ is the true anomaly at time $t_i$.

We can then produce an additional family of models, whose predictions at time $t_i$ are:
$$
m^{k(n)d(m)}_i = m^{k0d(m)}_i + \sum^n_{j=1}\vkep^{(j)}\;\;,
$$
where $n$ is the number of orbiting companions assumed, and $m$ is the degree of the polynomial assumed for the additional long-term velocity term.

For the error terms $\epsilon_i$ we explored two possibilities. In the first, the error terms are distributed according to a zero-centred multivariate normal with a covariance matrix whose elements are $\Sigma^{(WN)}_{ij} = \left(\sigma_i^2 + \sigma_J^2\right)\;\delta_{ij}$, where $\sigma_i$ is the internal uncertainty of the measurement at time $t_i$, $\sigma_J$ is a nuisance parameter called jitter, and $\delta_{ij}$ is the Kronecker delta. In other words, we assume the errors are independent. We call this the white noise (WN) model. Of course, this is strictly not true. Unaccounted-for effects, such as the rotational modulation of the star will translate to correlated errors in our model. Therefore, we also explored the possibility that the error terms are distributed according to a zero-centred multivariate normal with covariance function with elements given by:
$$
\Sigma^{(RN)}_{ij} = k_\mathrm{SE}(t_i, t_j) +  \delta_{ij}\;\left(\sigma_i^2 + \sigma_J^2\right)\;\;,
$$
where $k_\mathrm{SE}(t_i, t_j)$ is the squared exponential kernel function,
$$
k_\mathrm{SE}(t_i, t_j) = A^2 \exp\left(-\frac{(t_i - t_j)^2}{2\tau^2}\right)\;,
$$
with $A$ and $\tau$ the covariance amplitude and characteristic timescale, respectively. We call this hypothesis the red noise (RN) model. The chosen kernel function is general enough to describe a wide variety of temporal variations.

Under either of the two possibilities, the likelihood is a multivariate normal centred at the model prediction and with covariance $\Sigma$, whose natural logarithm is written
\begin{equation}
\log\mathcal{L}(\teta) = -\frac{1}{2} \left[ N\log(2\pi) + \log|\Sigma| + \left(\mathbf{v} - \mathbf{m}\right)^T \Sigma  \left(\mathbf{v} - \mathbf{m}\right) \right]
\label{eq.likelihood}
\end{equation}
where $\mathbf{v}$ and $\mathbf{m}$ are $N$-length vectors containing the velocity measurements and model predictions, respectively.

\subsection{Posterior distribution sampling}
\label{sect.mcmc}
The posterior distributions of the model parameters were sampled using the \emcee\ algorithm \citep{goodmanweare2010, emcee}, with 300 walkers. The prior distributions are listed in Table~\ref{table.priors}. For the parameters $\sqrt{e}\sin{i}$ and $\sqrt{e}\cos{i}$ we added an additional condition so that the eccentricity cannot acquire values above one. The algorithm was run for 25\,000 iterations for each model, and we discarded the first 5\,000 iterations for all models.  The chains exhibit adequate mixing and were checked for non-convergence. Summary statistics of the sampled posterior for individual models are presented in Tables~\ref{table.posteriorsk0wn} to \ref{table.posteriorsk1rn}. 

The stellar parameters are in excellent agreement between different models, except for the systemic velocity at Hiparccos epoch, $\gamma_0$, which changes slightly between models with different degree of polynomial acceleration, as expected. The posterior distributions of the stellar proper motion and parallax are dominated by the priors provided by the Hipparcos data. 

\begin{figure}
\input{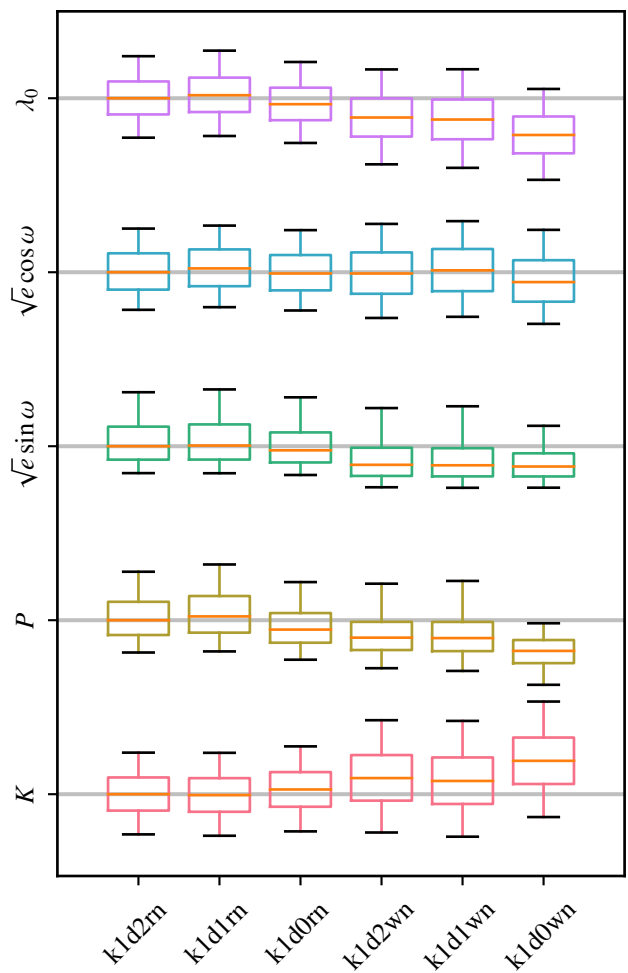}
\caption{Box plots for Keplerian parameter under all models tested. The boxes extend from the lower to upper quartiles, and the whiskers extend between the 5th and 95th percentiles. The red horizontal lines mark the position of the median. The grey horizontal lines have been chosen conveniently at the median values of the k1d2rn model.}
\label{fig.boxplots}
\end{figure}

The parameters of the Keplerian curve are also in good agreement across models, but the WN models produce a slightly larger amplitude and shorter period (Fig.~\ref{fig.boxplots}). The measured maximum-a-posteriori (MAP)-estimates of the RV amplitude range between 1.70 \ms and 1.91 \ms --the mean is slightly smaller (see Table~\ref{table.merged}). The MAP-estimates of the orbital period range from 12.9439 days to 12.9477 days. 

Concerning the acceleration terms, models with at least a constant additional acceleration are favoured by the data, as shown in Sect.~\ref{sect.modelcomp}. The amplitude of the constant acceleration, $\gamma_1$, depends only slightly on whether a higher degree polynomial is used to describe the data. The MAP-estimates for $\gamma_1$ under RN models range between -0.49 \ms/yr and -0.37 \ms/yr, which roughly implies companions with periods larger than around 13 years (twice the observation time span), and with minimum masses above 0.1 Jupiter masses.

The covariance parameters change significantly between models with and without a Keplerian curve. Besides the obvious fact that the additional white noise term is much larger in WN models than in RN models, RN model without a Keplerian curve exhibit a shorter covariance timescale --MAP estimates around 3.8 days for k0 models, and between 8.2 and 9.0 days for k1 models. This is reasonable: the models without a Keplerian curve require the increased flexibility provided by a shorter timescale \citep{rasmussenwilliams2005} to explain the full variability present in the time series. In the same sense, models without a Keplerian component also result in slightly larger covariance and white noise amplitudes.

To test the effects of moonlight contamination, we repeated the posterior sampling for the k1d1rn model, but removing either the four observations taken with the Moon closer than 30 degrees, or the 16 measurements acquired with the Moon closer than 35 degrees from Gl411. In both cases, the results are in excellent agreement with those using the full dataset. We therefore confirm that moonlight contamination is not relevant for this bright target.

\subsection{Periodogram analysis \label{sect.periodogram}}

\begin{figure}
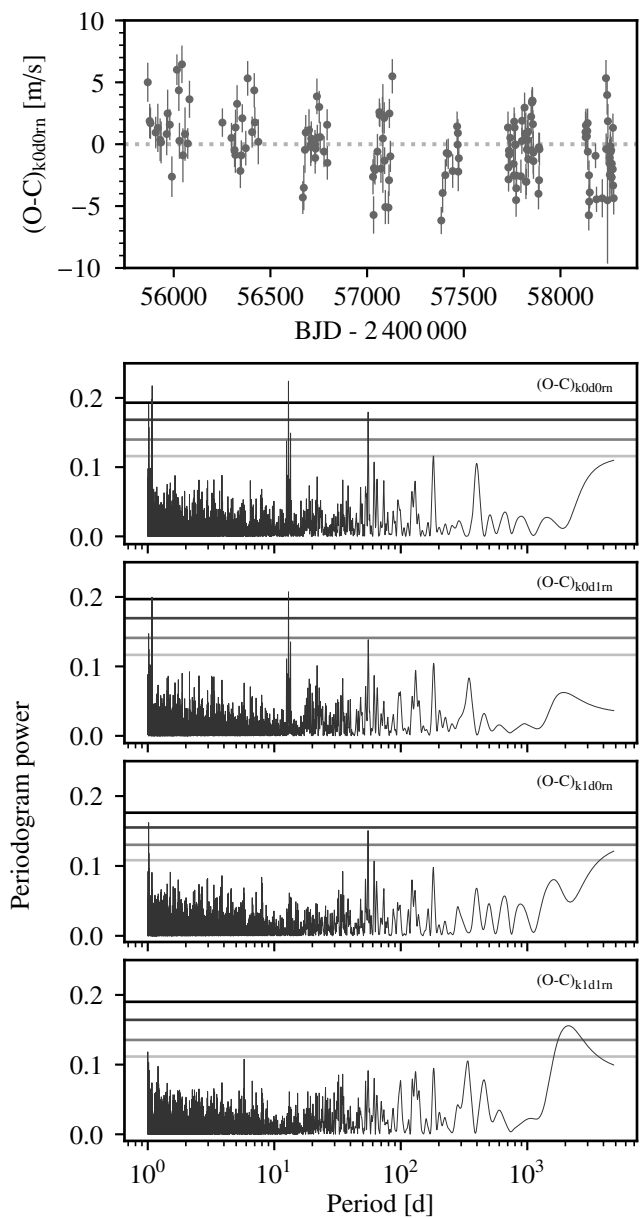

\flushleft{
\input{figures/Gl411_oc_k0d0rn.pgf}
\input{figures/Gl411_GLS.pgf}}
\caption{Radial velocity residuals of the MAP-estimate k0d0rn model (\emph{top panel}), and generalised Lomb Scargle periodograms under the red noise models. The horizontal lines in the periodogram panels correspond to p-value levels of 0.5, 0.1, 0.01, and $1\times10^{-3}$, from bottom to top, and are computed for each dataset by bootstrapping (see text for details). \emph{Second from top}: GLS of radial velocities after zero-point correction and removal of secular acceleration term, that is the residuals of the k0d0 model shown in the top panel; \emph{third from top}: residuals of k0d1 model, containing an additional constant acceleration; \emph{second from bottom}: residuals of the k1d0 model, with a Keplerian curve of period $P=12.9438$ days; \emph{bottom}: as above, but including also a linear acceleration term in the model (model k1d1). A version of the periodograms using a linear frequency scale in the $x$-axis is provided in Appendix~\ref{appendix.freqperiodograms} (Fig.~\ref{fig.periodogramsfreq}).}
\label{fig.periodograms}
\end{figure}

Fig.~\ref{fig.periodograms} presents a series of GLS periodograms of the RV residuals of models of increasing complexity. In every case, the MAP model is subtracted from the RV data presented in Table~\ref{table.rv}, and the MAP estimate of the additional white noise term, $\sigma_J$ was added in quadrature to the velocity uncertainties. The MAP parameters of each model are listed in Tables~\ref{table.posteriorsk0rn} and ~\ref{table.posteriorsk1rn}. In the residuals of the model including only the zero-point correction and a secular acceleration term (k0d0; top periodogram in Fig.~\ref{fig.periodograms}) powerful peaks are seen at $P = 12.945$ days and at the frequencies corresponding to the one-day aliases (aliases of one solar day and one sidereal day, 1.0837 d and 1.08048 d, respectively). No significance difference is seen between the white and red noise models at this point. Including a linear acceleration term (k0d1) does not reduce the peak powers significantly, as seen in the middle panel of Fig.~\ref{fig.periodograms}, but slightly changes the frequency at which the maximum occurs to $P = 12.959$ days.

The peak-to-peak variations of the instrumental zero-point are of the order of 15 \ms, and mask the presence of the planetary candidate signal (upper right panel of Fig.~\ref{fig.modelintime}). In the lower right panel of Fig~\ref{fig.modelintime} we show the GLS of the zero-point offset, interpolated at the epoch of observations, which does not show power at the period of the putative signal. It is therefore unlikely that the variability at the $P = 12.96$ days is introduced by either the zero-point correction or by a coupling of the low-frequency variations and the window function. We nevertheless explore this possibility in much more detail in Sect.~\ref{sect.sampling}.

The largest periodogram power is commonly used as a test statistic to evaluate the significance of the maximum peak in the periodogram. We estimate the corresponding $p$-values\footnote{The $p$-value of a measured peak power is usually called False Alarm Probability by the exoplanet community.} by shuffling the residuals 10\,000 times (solid horizontal lines in Fig~\ref{fig.periodograms}) and computing the maximum peak power in each realisation. The full distribution is estimated by means of a Gaussian kernel density method\footnote{The routine coded in the \texttt{scipy} package was used, with Scott's and Silverman's rule for the choice of the bandwidth.The results are independent of the bandwidth method used.}, using the resulting powers as input. Under the red noise models, the $p$-value of the observed maximum peak is estimated to be $9\times10^{-5}$ under the k0d0 model and $3\times10^{-4}$, under the k0d1 model. These values are low enough to warrant further investigation. 

The third periodogram in Fig.~\ref{fig.periodograms} represents the GLS of the residuals of a model including a Keplerian curve (k1d0), with a period around 12.94 days (see Table~\ref{table.posteriorsk1rn} for its MAP estimate). Neither of the peaks are present in this figure, and no other peak is seen with power above the 1\% $p$-value level. Some residual power is seen around 55 days, probably related to stellar activity and in agreement with the adopted rotational period, $P_\mathrm{rot} = 56.15\pm0.27$ days, but also possibly due to another planetary companion. However, this peak disappears when a model with a Keplerian curve and a linear acceleration term is considered (k1d1; bottom panel). A long-term periodicity ($\sim 2000$ days) is seen in the bottom panel, but with peak power below the 1\% $p$-value level.

\begin{figure*}
\includegraphics[width=\textwidth]{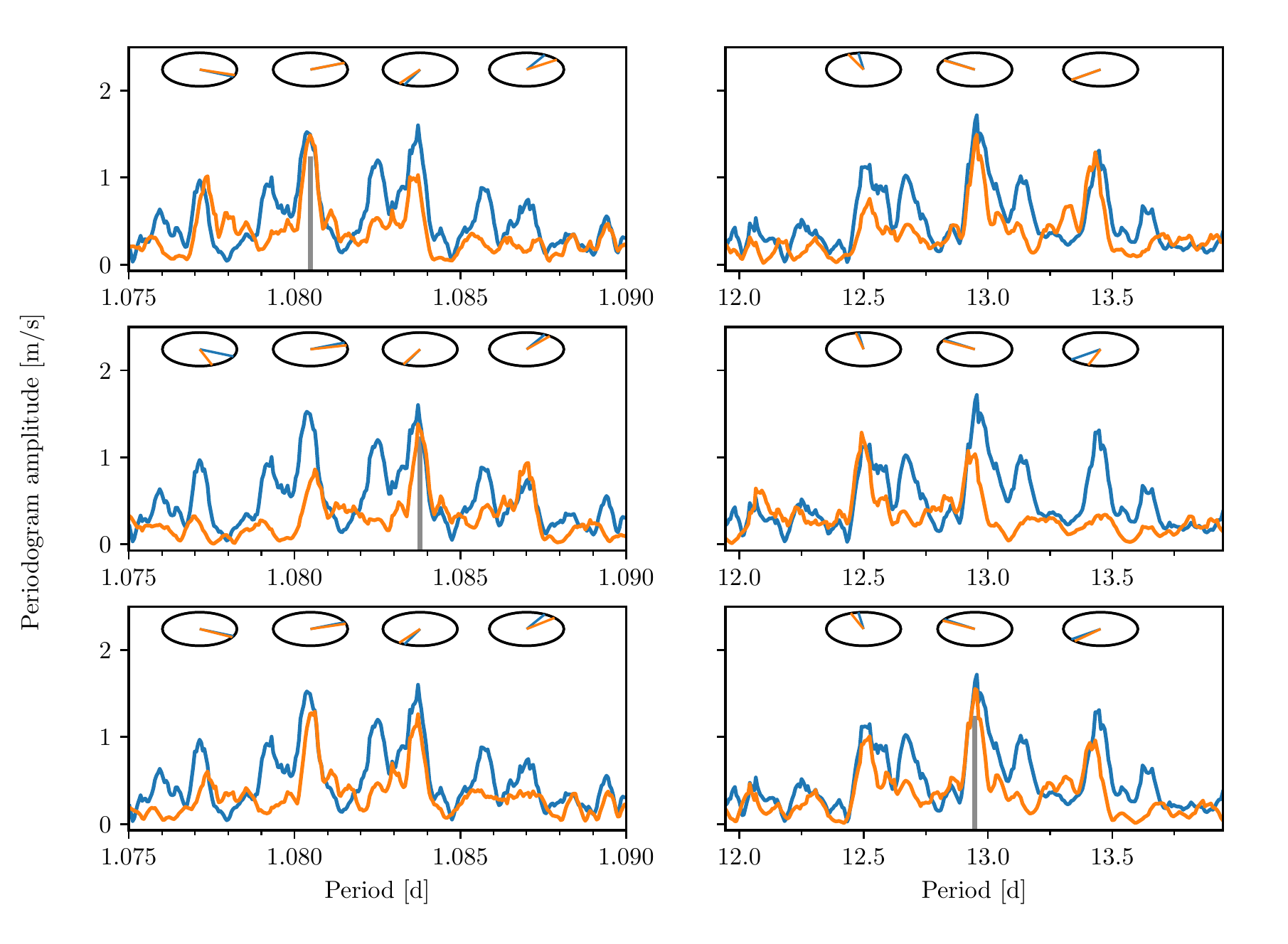}
\caption{Aliases evaluation. Peak amplitudes and phases in two regions of period space related by daily aliases (left and right columns). Each row contains simulations assuming a given sinusoid period (marked as a grey vertical line in each row) is present in the data. The dials on top of selected peaks indicate the phase at that period. The orange curves and dials correspond to the simulations, while the blue curves and dials are the periodograms computed on the real data, and are identical in all three rows. The simulated data containing a sinusoid at 1.084-d period does not reproduce the amplitudes of the peaks at 1.0805 d and 12.95 d. This period is therefore discarded.} 
\label{fig.aliases}
\end{figure*}

Although the peak at around $P = 12.95$ days is the strongest one, its 1-day aliases have similar power. The procedure described by \citet{dawsonfabrycky2010} was employed in an attempt to identify the real periodicity. The method consists in simulating datasets using the time stamps of the real data, and injecting a sinusoidal signal at each of the candidate periodicities. The amplitudes and phases of the peaks in the periodogram of the resulting synthetic data are compared to the actual observations. The results are shown in Fig.~\ref{fig.aliases}, where each row corresponds to a different tested periodicity, indicated by a vertical grey line. The two relevant regions of the periodograms are presented. The blue periodograms are computed for the actual data, and are identical in all three rows. The orange one is computed on each simulated data set. On top of each relevant peak, we present a dial representing the phase of the signal. It seems clear that the peak at 1.0837 d is not a real one; on the other hand, some doubt still remains between the peak at 12.9478 d and the one at 1.0805 d. The former case reproduces the amplitude of the aliases structure and its phases correctly, in particular the peaks around 1.0835 d and 12.5 d, which are not correctly reproduced with the simulated sinusoid at 1.0805 d. On the other hand, the simulation with a sinusoid at 12.9478 d does not reproduce the peak observed at 1.077 d. The phases do not seem to provide much additional information in this case. In light of this, we assume the real period of the signal to be around $P=12.9478$ days, but warn that the other option cannot be fully discarded at this time. In Sect.~\ref{sect.conclusion} we present further evidence for the 12.95-day period. In any case, additional high-cadence observations should permit distinguishing between both periods more precisely.

\subsubsection*{Ancillary observables}

Here we study the variability of selected ancillary observables, the bisector velocity span, the FWHM of the CCF, and the activity proxies based on the H$\alpha$ and Ca II H \& K line fluxes. The FWHM and bisector span exhibit a long-term variation that is seen in other stars of the SP3 sample as well. Therefore, we corrected these observables from this effect by subtracting a quadratic fit from the time series. 

\begin{figure*}
\input{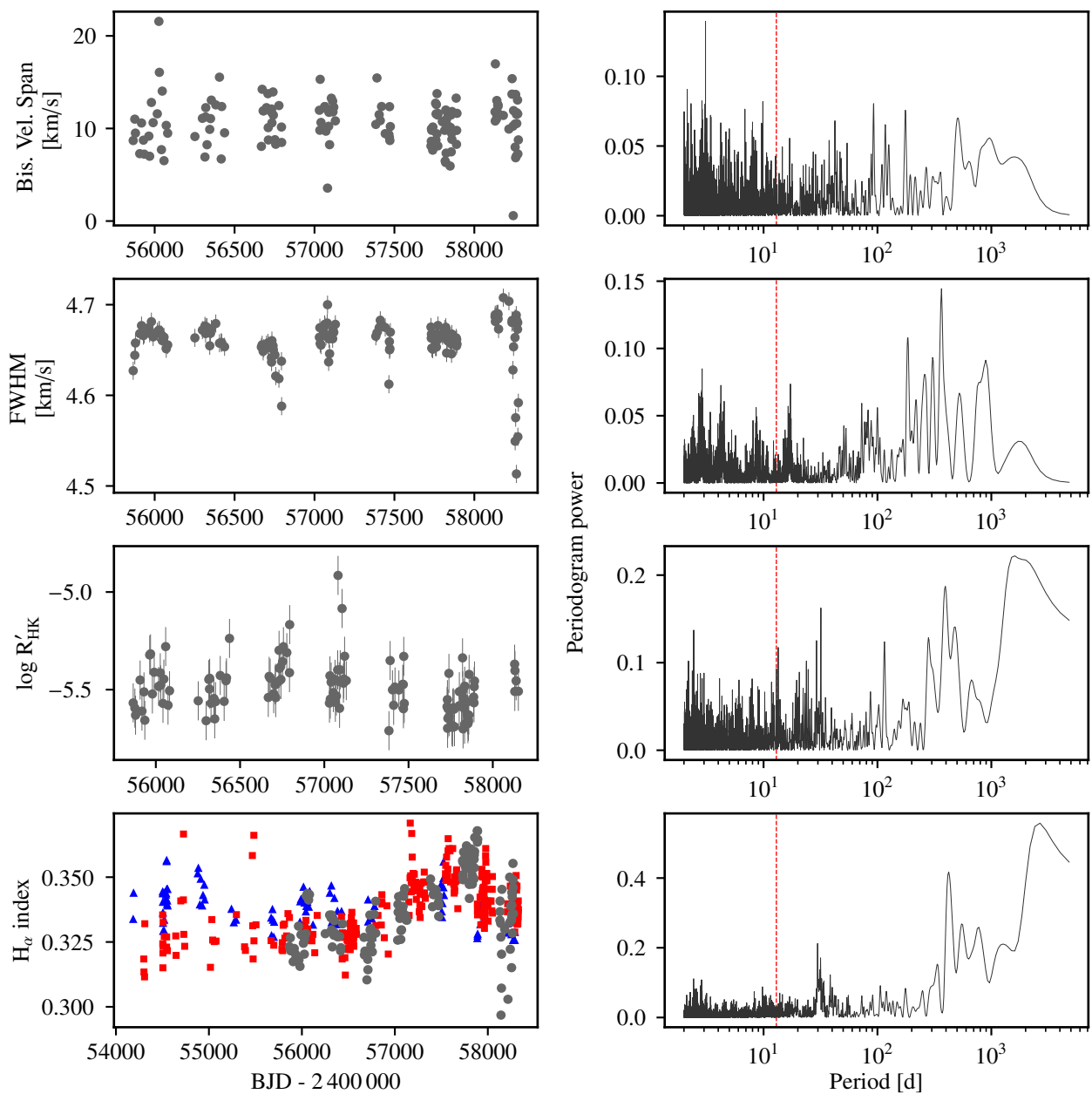}
\caption{Time series (\emph{left}) and GLS periodograms (\emph{right}) of selected ancillary observables. The best-fit second order polynomials were subtracted from the Bisector Velocity Span and the FWHM time series. The red vertical lines in the periodograms indicate the frequency of the detected planet candidate. In the H$\alpha$ time series, we also present the data corresponding to two other SP3 targets (see text for details): blue triangles for Gl514 and red squares for Gl686. A version of the periodograms using a linear frequency scale in the $x$-axis is provided in Appendix~\ref{appendix.freqperiodograms} (Fig.~\ref{fig.GLSancifreq}).}
\label{fig.GLSanci}
\end{figure*}

In Fig.~\ref{fig.GLSanci} we present the time series and GLS periodograms of these ancillary observables. None of the periodograms exhibit power at the frequency detected in the radial velocity time series. No variation is seen either at the expected rotational period reported in Table~\ref{table.stellarparams}, around 56 days, nor at the periods of the other estimated rotational periods discussed above. On the other hand, the activity indices based on the Ca II (\logR) and H$_\alpha$ lines exhibit some power at periods around 31 days, but with large $p$-values. Peaks close to one year are also seen in the FWHM time series as well as in the activity proxies, indicating some sort of systematic error still present in SOPHIE data, or an incomplete treatment of the telluric line contamination. In the case of the FWHM, this variability could also be a residual of the window function, due to an incorrect correction for the long-term variation described above. When the 1-year period is subtracted from the data, no other significant periodicities are detected.

Finally, a long-term activity variation is seen, specially in the H$\alpha$ index, with a maximum in the second to last season. However, other stars observed intensively by the SP3 exhibit a similar increase in the H$\alpha$ level at this time. In  the bottom panel of Fig~\ref{fig.GLSanci} we also present the time series for Gl686 (red squares) and Gl514 (blue triangles), two standard stars used in the SP3, offset to the mean value of the Gl411 time series. The increase in the H$\alpha$ level between JD=2\,457\,000 and JD=2\,458\,000 is seen in all three stars and point to a not-yet-understood systematic effect. We tried to perform a similar analysis for the \logR\ time series, but the noise in the \logR\ time-series of these stars is too large to allow us to draw meaningful conclusions. We therefore do not claim detecting a long-term activity variation.

The ancillary observations therefore give no reason to confer an activity or systematic origin to the radial velocity peak at $P\sim12.9$ days.

\subsection{Model comparison \label{sect.modelcomp}}
The periodogram analysis of Sect.~\ref{sect.periodogram} suggests a periodic component is present in the data with a period of 12.945 days. However, the presence of strong power at this frequency does not provide definitive evidence for the 1-Keplerian model\footnote{For a discussion of common misinterpretations of the p-values and associated issues, see, for example, \citet{sellke2001}.}. Instead, the Bayesian approach allows us to compute the posterior probability of different models and to compare them directly. In particular, the posterior odds ratio between two competing models reduces to the ratio of their marginal likelihoods if equal prior probabilities are assumed. 

We used the posterior samples from different models to estimate the corresponding marginal likelihoods via the importance sampling estimator introduced by \citet{perrakis2014}, as implemented by \citet{diaz2016a}. This estimate was shown to provide results in agreement with a number of other estimators of the marginal likelihood \citep{nelson2018}. We further compared the results from preliminary computations with the models employed in this paper using the nested sampling algorithm Polychord \citep{handley2015}. The uncertainty in this determination was estimated through a MonteCarlo approach repeating the computation 100 times. More details on our implementation of the algorithm are given in Appendix A.6 of \citet{nelson2018}. 

\begin{figure}
\input{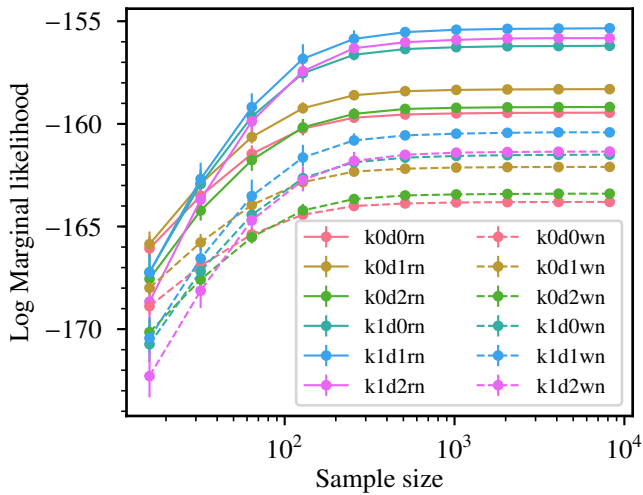}
\caption{Estimation of the marginal likelihood as a function of the sample size used to compute it. Dashed lines correspond to white noise models, continuous lines to red noise models. The estimator is defined in \citet{perrakis2014}, and the details of the implemention are given in \citet{diaz2016a, bonfils2018} and in the Appendix of \citet{nelson2018}.}
\label{fig.evidences}
\end{figure}

Although the estimator is known to be biased \citep{perrakis2014}, it was shown previously that the bias is negligible for sample sizes larger than a few thousands \citep[e.g.][]{bonfils2018, hobson2018}. We verified this by computing the estimator of the marginal likelihood for an increasing number of samples from the joint posterior (Fig.~\ref{fig.evidences}). We reach the same conclusion as in previous works. For sample sizes above 1000 or 2000 the bias does not seem significant. The results based on a sample of size 8192 are reported in Table~\ref{table.evidences}. 

Overall, it is seen that models containing a Keplerian curve are favoured over models without a Keplerian component. The evidence for models with one Keplerian is much stronger under the red noise models, with Bayes factors above 900. Under the white noise hypothesis, Bayes factors for the k1d1 and k1d0 models are around 48 and 4, respectively, when compared with model k0d1. These are still considered as "strong" and "positive" evidence, respectively, against models without Keplerians \citep{kassraftery1995}. In any case, models accounting for correlation in the data (red noise models) are largely favoured over white noise models, with Bayes factors exceeding $10^5$ for models with one Keplerian. Finally, this computation shows that the data are not sufficient to distinguish between models k1d0, k1d1, and k1d2, but provide a slight preference for the model with a constant acceleration (k1d1). The posterior marginal distributions are similar under all three models with a Keplerian component. 

In view of these results, we decided to produce a sample of the parameter posterior distribution averaged over models, as we did in \citet{bonfils2018}. For this, we resampled the posterior samples obtained for each individual model using the Bayes factors as weights. In Table~\ref{table.merged} we provide summary statistics of the marginal distributions of the model parameters and a few additional quantities of interest. In this manner, the Keplerian parameters reported are independent of whether the additional acceleration exists or not, or on the chosen noise model, although in practice, given the strong support for the red noise model, no samples from the white noise models are present in the final merged posterior sample.

\begin{figure}
\center
\input{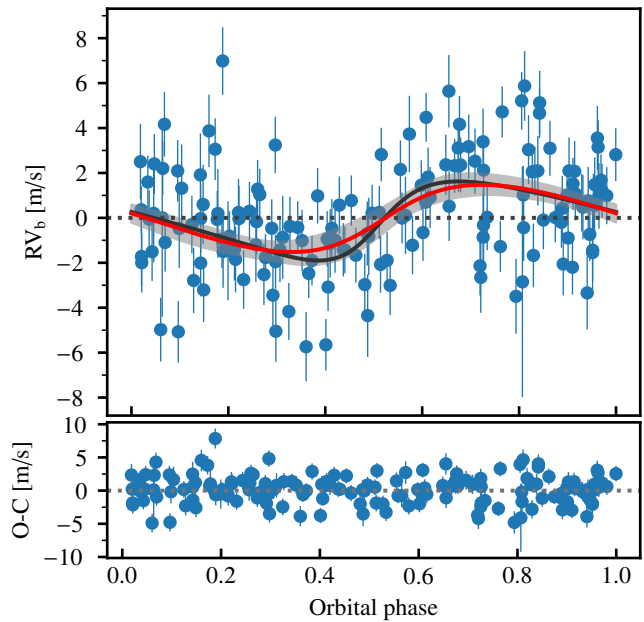}
\caption{SOPHIE velocities, after subtraction of MAP-estimate model for the zero-point offset, secular acceleration and additional acceleration terms, and phase folded to the MAP-estimate of the Keplerian period, under the k1d1 hypothesis with correlated noise. The black and red curves represent the MAP-estimate and mean Keplerian model over the posterior samples obtained with the MCMC algorithm. The shaded region extends between the 5th- and 95th-percentile model curves, computed over 10\,000 randomly chosen posterior samples.}
\label{fig.phasefolded}
\end{figure}

In Fig. \ref{fig.modelintime} we present the slowly-varying components of all three models, that is the components remaining after removing the Keplerian contribution. Fig.~\ref{fig.phasefolded} represents the SOPHIE radial velocities, corrected from the MAP-estimate of slowly-varying terms in the model and phase-folded to the MAP period estimate, together with the Keplerian curve (MAP and mean) and its confidence intervals under the most probable hypothesis (k1d1, with correlated noise). The dispersion of the residuals of the MAP model is 2.18 \ms, which according to our model is produced in almost equal parts by the photon noise plus calibration error, whose mean value is 1.4 \ms, and the covariance amplitude, whose MAP value is 1.85 \ms.

\begin{table*}[t]
\centering
\caption{Log Marginal likelihood ($\evid$) for different models, and Bayes Factor with respect to model k1d1 for SOPHIE data.} \label{table.evidences}

\begin{tabular}{l c c c c c c}        
\hline\hline                 
& \multicolumn{2}{c}{White Noise}& \multicolumn{2}{c}{Red Noise}\\
Model & $\log_{10}{\evid}$ & $\log_{10}{\mathrm{BF}_{\mathrm{k1d1}, \mathbf{\cdot}}}$ & $\log_{10}{\evid}$ & $\log_{10}{\mathrm{BF}_{\mathrm{k1d1}, \mathbf{\cdot}}}$ \\
\hline
k0d0 & $-163.8024 \pm 0.0017$ & $3.390 \pm 0.019$ &$-159.4534 \pm 0.0068$&$4.113 \pm 0.077$\\	
k0d1 & $-162.0948 \pm 0.0019$ & $1.683 \pm 0.019$ &$-158.3071 \pm 0.0085$&$2.966 \pm 0.077$\\
k0d2 & $-163.400 \pm 0.012$     & $2.988 \pm 0.022$ &$-159.177  \pm  0.016$  &$3.836 \pm 0.080$\\
k1d0 & $-161.498 \pm 0.031$ & $1.086 \pm 0.036$ &$-156.197 \pm 0.023$&$0.856 \pm 0.080$\\
k1d1 & $-160.412 \pm 0.019$ & 0.0 	&$-155.341 \pm 0.077$&0.0	            \\
k1d2 & $-161.347 \pm 0.036$ & $0.935 \pm 0.041$ &$-155.823 \pm 0.034$&$0.482 \pm 0.085$\\
\hline
\end{tabular}
\end{table*}

\begin{table*}
\centering
\caption{Model parameters averaged over tested models. Values correspond to the posterior sample mean, with errors being the standard deviation of the MCMC samples. In the second line we report the 95\% highest density interval (HDI).}
\label{table.merged}
\begin{tabular}{llc}
\hline
\hline\noalign{\smallskip}
\hline\noalign{\smallskip}
\multicolumn{3}{l}{\bf Stellar parameters}\\ 
\hline\noalign{\smallskip}
Systemic velocity\tablefootmark{$\dagger$}, $\gamma_0$ & [km/s] & $-84.64566 \pm 0.00062$  \\ 
&& $[-84.64712, -84.64467]$ \\ 
\noalign{\vspace{0.75mm}}
Parallax\tablefootmark{$\dagger$}, $\pi$ & [mas] & $392.64 \pm 0.67$  \\ 
&& $[391.32, 393.97]$ \\ 
\noalign{\vspace{0.75mm}}
Proper motion (RA)\tablefootmark{$\dagger$}, $\mu_\alpha$ & [mas/yr] & $-580.27 \pm 0.62$  \\ 
&& $[-581.50, -578.97]$ \\ 
\noalign{\vspace{0.75mm}}
Proper motion (DEC)\tablefootmark{$\dagger$}, $\mu_\delta$ & [mas/yr] & $-4765.85 \pm 0.64$  \\ 
&& $[-4767.21, -4764.61]$ \\ 
\noalign{\vspace{0.75mm}}
\hline\noalign{\smallskip}
\multicolumn{3}{l}{\bf Planet parameters}\\ 
\hline\noalign{\smallskip}
Orbital period, $P$ & [d] & $12.9532 \pm 0.0079$  \\ 
&& $[12.9388, 12.9697]$ \\ 
\noalign{\vspace{0.75mm}}
RV amplitude, $K$ & [m/s] & $1.59 \pm 0.23$  \\ 
&& $[1.14, 2.09]$ \\ 
\noalign{\vspace{0.75mm}}
Orbital eccentricity, $e$ & [0..1[ & $0.22 \pm 0.13$  \\ 
&& $[0.00, 0.44]$ \\ 
\noalign{\vspace{0.75mm}}
Mean longitude, $\lambda_0$ & [deg] & $-39.1 \pm 9.4$  \\ 
&& $[-57.3, -19.7]$ \\ 
\noalign{\vspace{0.75mm}}
\noalign{\vspace{0.75mm}}
Minimum mass, $M_p \sin i$ & [$\mathrm{M}_\oplus$] & $2.99 \pm 0.46$  \\ 
&& $[2.08, 3.95]$ \\ 
\noalign{\vspace{0.75mm}}
Semi-major axis, $a$ & [AU] & $0.0785 \pm 0.0027$  \\ 
&& $[0.0732, 0.0837]$ \\ 
\noalign{\vspace{0.75mm}}
Eq. temperature (albedo=0.3) & [K] & $349.83 \pm 0.32$  \\ 
&& $[349.44, 350.45]$ \\ 
\noalign{\vspace{0.75mm}}
Eq. temperature (albedo=0.8) & [K] & $255.77 \pm 0.24$  \\ 
&& $[255.48, 256.22]$ \\ 
\noalign{\vspace{0.75mm}}
$\sqrt{e} \cos{\omega}$ &  & $-0.07 \pm 0.23$  \\ 
&& $[-0.53, 0.38]$ \\ 
\noalign{\vspace{0.75mm}}
$\sqrt{e} \sin{\omega}$ &  & $-0.31 \pm 0.24$  \\ 
&& $[-0.70, 0.19]$ \\ 
\noalign{\vspace{0.75mm}}
Time of inferior conjuction, $T_c$ & [BJD] & $57184.51 \pm 0.43$  \\ 
&& $[57183.58, 57185.29]$ \\ 
\noalign{\vspace{0.75mm}}
Distance at inferior conjuction, $r_{Tc}$ & [AU] & $0.065 \pm 0.011$  \\ 
&& $[0.042, 0.086]$ \\ 
\noalign{\vspace{0.75mm}}
Transit probability &  & $0.0289 \pm 0.0055$  \\ 
&& $[0.0192, 0.0400]$ \\ 
\noalign{\vspace{0.75mm}}
\hline\noalign{\smallskip}
\multicolumn{3}{l}{\bf Drift parameters\tablefootmark{$\ddagger$}}\\ 
\hline\noalign{\smallskip}
Additional acceleration, $\gamma_1$ & [m/s/year] & $-0.49 \pm 0.16$  \\ 
&& $[-0.84, -0.16]$ \\ 
\noalign{\vspace{0.75mm}}
Quadratic acceleration, $\gamma_2$ & [(m/s/year)$^2$] & $0.245 \pm 0.099$  \\ 
&& $[0.049, 0.454]$ \\ 
\noalign{\vspace{0.75mm}}
\hline\noalign{\smallskip}
\multicolumn{3}{l}{\bf Noise model parameters}\\ 
\hline\noalign{\smallskip}
Additional white noise term, $\sigma_J$ & [m/s] & $0.45 \pm 0.28$  \\ 
&& $[0.00, 0.95]$ \\ 
\noalign{\vspace{0.75mm}}
Covariance amplitude, $A$ & [m/s] & $1.85 \pm 0.29$  \\ 
&& $[1.25, 2.45]$ \\ 
\noalign{\vspace{0.75mm}}
Characteristic timescale, $\ln\tau$ & [d] & $2.21 \pm 0.25$  \\ 
&& $[1.66, 2.70]$ \\ 
\noalign{\vspace{0.75mm}}
Zero-point correction constant, $a_c$ & -- & $0.74 \pm 0.12$  \\ 
&& $[0.49, 0.97]$ \\ 
\noalign{\vspace{0.75mm}}
\hline
\end{tabular}
\tablefoot{
\tablefootmark{$\dagger$}Measured at Hipparcos epoch.
\tablefootmark{$\ddagger$}Computed only over models including additional drift terms.}.

\end{table*}

\subsection{Validation of planetary model}

The detected radial velocity signal has the smallest-amplitude candidate signal detected by the SOPHIE spectrograph so far. We are therefore inclined to be particularly cautious about its origin. In this Section we present a series of tests to validate its planetary origin.

\subsubsection{Stacked periodograms}
In Sect.~\ref{sect.periodogram} we showed that none of the activity proxies present clear variability at periods related to the planet candidate signal. Another useful diagnostic for activity is the stacked periodograms, which consists in studying the evolution of the power of a periodogram peak as the number of observations in the time series is increased \citep[for details, see][]{mortiercameron2017}. A planetary signal should increase its power monotonically, within the limits of the noise, with the number of observations, given that a proper normalisation for the noise level under different sample sizes is taken into account.

\begin{figure}
\center
\input{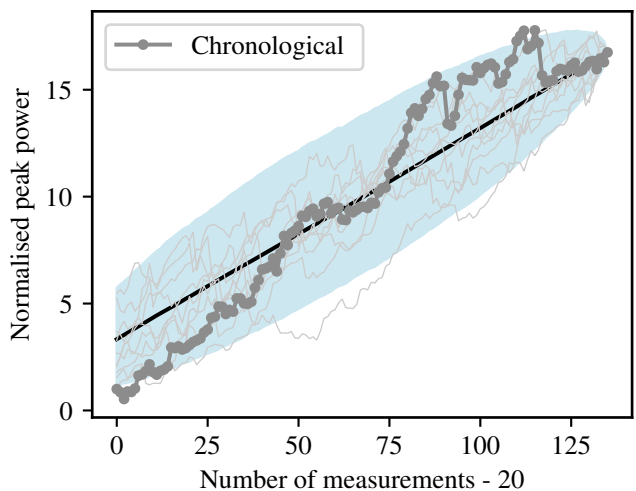}
\caption{Evolution of the normalised power of the periodogram peak around the planetary candidate period. The GLS periodogram is normalised assuming Gaussian noise remains after removing the signal \citep[Eq.~22 in][]{zechmeisterkurster2009}. The grey curve and points correspond to the case where the measurements are added chronologically, while the light blue shaded area encompasses the range between the 5th- and 95th-percentile of the distribution obtained when the data are included in random order. The black curve is the mean of the distribution, and the thin curves are ten draws from the distribution.}
\label{fig.stacked}
\end{figure}

In Fig.~\ref{fig.stacked} we present the evolution as the number of data points considered increases of the peak power computed on the residuals of model k0d1, with correlated noise, and normalised according to Eq. 22 in \citet{zechmeisterkurster2009}. The maximum peak between 12.9 and 13.0 days was considered, to allow for the possibility that the peak position may vary as new data are included. The data points were added chronologically (grey curve and points) or randomly. For the latter case, we produced 10\,000 realisations for each number of data points\footnote{There are over 10'000 possible combinations of $n$ data points chosen out of the 155 available velocity points for all $n<154$.}, changing randomly the order in which the data points were included in the analysed set. The shaded light blue area encompasses the area between the 5th- and 95th-pecentile of the obtained distribution, and the black curve is the mean of the distribution. The normalised peak power increases with the number of points considered, lending support to the hypothesis that the variation is produced by an authentic periodic signal. Additionally, adding the measurements chronologically produces a curve with similar features as the ones obtained by randomly shuffling the data order (thin grey curves), supporting the idea that the signal is not produced by an effect --systematic or activity-- occurring at a given moment in the time series.

\subsubsection{Possible sampling artefact \label{sect.sampling}}

Periodicities introduced by the observational sampling can produce spurious detections in radial velocity data \citep[e.g.][]{rajpaul2015}. In particular, as our model contains low-frequency terms, one may worry that if they are incorrectly accounted for, the window function \citep{roberts1987} will be imprinted in the RV time series, producing spurious signals. In this section we explore if this effect can produce a signal at the frequency of the putative companion.

\begin{figure}
\input{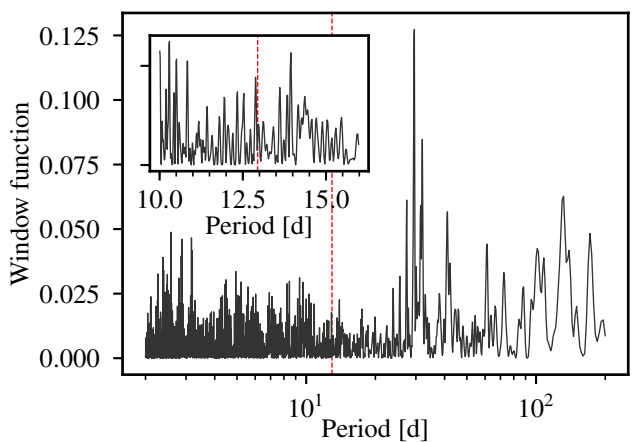}
\caption{Window function of the SOPHIE observations for periods between 2 d and 200 d. The largest peaks in the main panel correspond to the Moon synodic period. In the inset panel we present a zoom around the MAP planet period, indicated in both axes by a vertical dashed line.}
\label{fig.windowfunction}
\end{figure}

As a first step, we studied the window function in detail. Besides the strong peaks at and around 1-day and 1-sidereal day, the largest power at periods shorter than 200 days is at a group of peaks around 29.55 days, the Moon synodic period\footnote{Because of their relative faintness, SP3 targets are usually avoided around the full moon. Gl411 is too bright and has a velocity too far from the Moon to be affected, but as this target was considered as a standard star for the SP3, it was not observed when observations of the programme were not performed.} (see Fig.~\ref{fig.windowfunction}). The region around the frequency corresponding to the planetary candidate is devoid of strong power. It is therefore unlikely that the low-frequency terms induce a signal at the candidate period.

\subsubsection*{White noise simulations\label{sect.wnsimul}}
To verify this, we generated synthetic data sets sampled at the observing dates of the real SOPHIE observations. The synthetic data contain the effect of the secular acceleration and zero-point correction, but no planet or additional long-term velocity variations were included. The parameters of these effects were randomly drawn from the joint posterior sample of the k1d1rn model, obtained in Sect.~\ref{sect.mcmc}. We added random white noise with a variance equal to the sum of the internal observational error and the inferred jitter (white and correlated), 
$\sigma_i^2 + \sigma_J^2 + A^2$. We generated in this manner 10\,000 synthetic radial velocity time series. For each, we fitted the parameters of the k0d0 model using a variation of the Levenberg-Marquardt algorithm implemented in {\tt scipy}, starting at the MAP-estimate. A GLS was performed on the residuals of the fit, scanning periods between 2 and 200 days, to avoid the obvious peaks coming from the window function. This is similar to what was done to obtain the periodogram in the upper panel of Fig.~\ref{fig.periodograms}. The maximum power within 0.1 days of the MAP period estimate, 12.947 days, is 0.14. This is to be compared to the value of 0.21 found in the real dataset. Additionally, the number of times the largest peak power appeared in this region was in agreement with what would be expected from white noise. Therefore, a combination of white noise and sampling cannot explain the observed peak at the candidate period. This is actually expected as the long-term variability can be correctly taken into account under these conditions \citep[see discussion in][]{rajpaul2015}.

\subsubsection*{Correlated noise simulations\label{sect.simulrn}}

The question remains whether correlated noise could produce a spurious detection. In this case, one may expect the low-frequency variations due to zero-point shifts and the secular acceleration to be masked by correlated noise, leading to an incorrect determination, and hence a residual of the sampling function present in the time series. To explore this, we repeated the experiment described above but including correlated Gaussian noise with a covariance produced by a kernel function of the quasi-periodic type:
$$
k_\mathrm{QP}(t_i, t_j) = A_\mathrm{act}^2 \exp\left(-\frac{(t_i - t_j)^2}{2\tau^2} - \frac{2}{\epsilon^2}\sin^2{\left(\frac{\pi (t_i - t_j)}{\mathcal{P}}\right)}\right)\;,
$$
where the four hyper-parameters are the amplitude of the covariance term ($A_\mathrm{act}$), the rotational period of the star ($\mathcal{P}$), the covariance decay time ($\tau$) and a shape parameter ($\epsilon$). This kernel is known to correctly represent the covariance produced by active regions rotating in and out of view \citep[e.g.][]{haywood2014, rajpaul2015}. To the covariance produced by this kernel, we added a diagonal matrix with the square of the internal data point uncertainties.

For each of the 10\,000 synthetic velocity time series produced, the values of the hyper-parameters were drawn from distributions which we claim represent fairly well our knowledge on Gl411, and the evolution of active regions in general. The parameter representing the rotational period, $\mathcal{P}$, was drawn from a normal distribution with mean 56.15 days, and variance $(0.27\;\mathrm{days})^2$, in agreement to our conclusion from Sect.~\ref{sect.stellar}. The typical lifetime of the active regions, represented by the parameter $\tau$ is $k\cdot\mathcal{P}$, where $k$ is randomly drawn from a $N(3, 0.2^2)$ distribution; the structure parameter $\struct$ is randomly chosen between 0.5 and 1. For the amplitude $A_\mathrm{act}$, we resorted to the posterior samples of the k0d1 model, for which we computed the quadratic sum of the covariance amplitude --with the squared-exponential kernel-- and the additional white noise term: $A^2 + \sigma_J^2$. Our \emph{ansatz} is that $A_\mathrm{act}$ is drawn from a normal distribution with mean and standard deviation taken from the covariance diagonal under the k0d1 model, $N(2.1\,\mathrm{ms}^{-1}; 0.0484\,\mathrm{m^2s^{-2}})$. In this manner, the simulations will have a similar dispersion around the k0 models as real data does.

\begin{figure}
\input{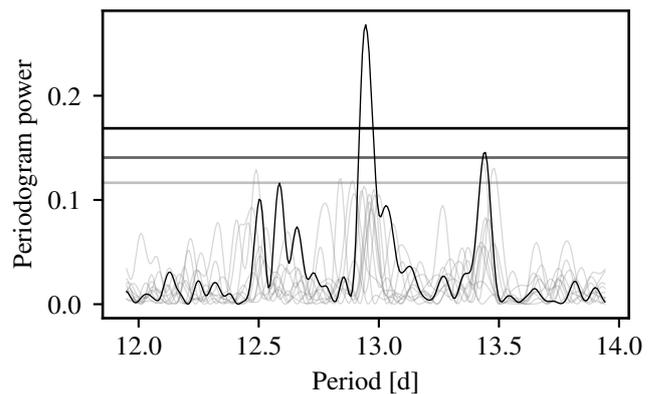}
\caption{Comparison of the actual GLS periodogram of SOPHIE data (dark curve) and the periodograms obtained from simulated data sets not including a signal at that period. The ten periodograms exhibiting the strongest power at the vicinity of the detected signal among the 10\,000 simulated cases are plotted in light grey. The horizontal lines are the 0.5, 0.1 and 0.01 $p$-value levels, computed as for Fig.~\ref{fig.periodograms}.}
\label{fig.simulrn_gls}
\end{figure}

None of the 10\,000 simulations exhibit power in the vicinity of the detected signal as strong as the one observed in the data (Fig.~\ref{fig.periodograms}). In Fig.~\ref{fig.simulrn_gls} the original GLS periodogram\footnote{To make a fair comparison, the additional noise term is not added to the uncertainties in this plot. This is the reason why the plot is not identical to the one presented in Fig.~\ref{fig.periodograms}} is presented together with the ten simulated periodograms exhibiting the strongest power within 0.1 days of the MAP-estimate period, 12.948 days. It can be seen that the simulated periodograms are not able to reproduce the observed peak.

\begin{figure}
\input{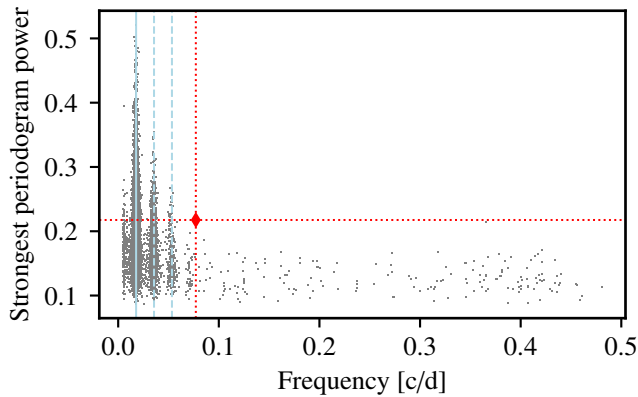}
\caption{Distribution of the power and frequency of the largest periodogram peaks for 10\,000 datasets simulated including only correlated noise. The rotational period and its two first harmonics are indicated by light blue solid and dashed lines, respectively. The red diamond represents the largest peak observed in the real dataset. 
}
\label{fig.simulrn_distribution}

\end{figure}

Additionally, following \citet{ribas2018}, the distribution of the frequency and power of the largest peaks in the 10\,000 simulated periodograms was studied (Fig.~\ref{fig.simulrn_distribution}). As expected, strong power is seen at and around the simulated rotational periods. This is not observed in the real data, suggesting an overestimation of the amplitude of simulated noise. None of the simulated periodograms exhibit the strongest peak at a frequency equal or greater (i.e. farther from the expected rotational signal) than the real peak with as much power. We can therefore provide a conservative upper limit for the $p$-value for the detected signal under the hypothesis of correlated noise of $1\times10^{-4}$. Finally, no peak was found --irrespective of whether it constitutes the largest peak in the simulated periodogram or not-- at frequencies higher than the detected one and with larger power.

We conclude that correlated noise is extremely unlikely to be the cause of the detected signal, which we therefore attribute to a planetary-mass companion in orbit around Gl411.

\section{Discussion \label{sect.discussion}}

The analysis from the previous section indicates that the detected signal is produced by a \emph{bona fide} planetary-mass companion. We discuss here the implications of this result.

\subsection{A temperate super-Earth around a nearby star. \label{sect.warm}}

\begin{figure}
\center
\input{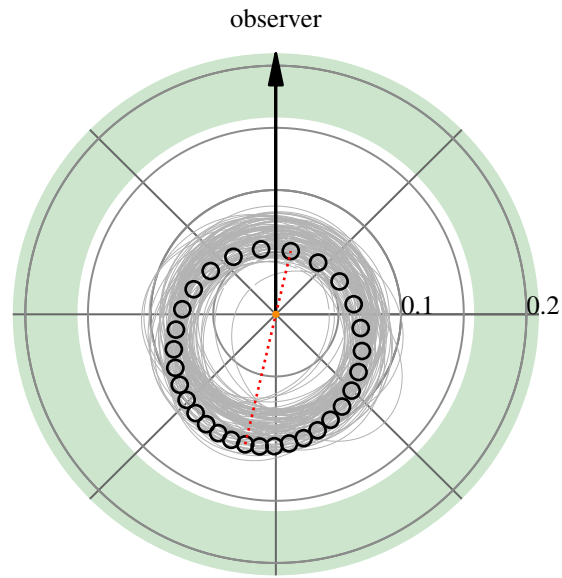}
\caption{Sketch of the orbit of Gl411 b, seen from above the orbital plane. The maximum-a-posteriori orbit is indicated with the empty black points that are equally spaced in time over the orbit, and the corresponding major axis is represented by the thin dotted red line. The thin grey lines are a hundred random samples from the merged posterior distributions (see text for details). The host star is at the centre, represented by an orange circle, whose size is to scale. The concentric circles are labelled in astronomical units and the black thick arrow points towards the observer. The filled green area is the inner region of the habitable zone, computed according to \citet{kopparapu2013}. The planet orbits the star too close to be habitable.}
\label{fig.orbitabove}
\end{figure}

Assuming the stellar parameters listed in Table~\ref{table.stellarparams}, the companion around Gl411, Gl411 b, has a minimum mass of $2.99\pm0.46\;\mathrm{M}_\oplus$, and the semi-major axis of its orbit is $0.0785\pm0.0027$ AU. This implies that the equilibrium temperatures assuming Earth-like and Venus-like Bond albedos (0.3 and 0.8, respectively) are around 350 K and 256 K, which are higher than required for the planet to be in the habitable zone \citep{kopparapu2013}. In Fig.~\ref{fig.orbitabove} we present a sketch representation of the orbit of Gl411 b and the location of the habitable zone \citep{kopparapu2013}. With the current data, a significant eccentricity is not detected. The MAP-estimate eccentricity is around 0.35, but the 95\%-HDI for the orbital eccentricity reported in Table~\ref{table.stellarparams} includes zero.

Being one of the brightest M-dwarf stars in the sky, Gl 411 is particularly prone for follow-up studies. Searching for transits of Gl 411 b is particularly relevant. Given the relatively small stellar radius, transits of Earth-size planets would produce a decrease of 0.05\% in the stellar flux. An educated guess of the planet radius of Gl411 b can be  obtained based on its mass, using the formalism from \citet{chenkipping2017}. This gives a radius for the planet of $R_p = 1.66^{+0.72}_{-0.42}\;R_\oplus$, where the minimum mass was multiplied by 1.15, to account for the unknown orbital inclination angle \citep{lovis2017}. The posterior transit probability\footnote{This posterior probability does not take into account the underlying mass distribution of exoplanets.} is of 2.9\%. At the present time, the transit time is known with a precision of over ten hours, and the width of the 95\%-HDI for the time of inferior conjunction is around 1.7 days. Therefore, transit search from the ground may be impractical. On the other hand, the Transiting Exoplanet Survey Satellite \citep[TESS;][]{ricker2014} and the Characterising Exoplanet Satellite \citep[CHEOPS;][]{broeg2013} could search for transits of the planetary companion down to Earth-size objects. However, the former will not observe Gl411 before late 2019, and the latter is currently scheduled to launch in 2019. 

Even in the absence of transits, the proximity of Gl411 to the Solar system makes its companion an ideal target for atmospheric characterisation. In Fig.~\ref{fig.teqcontext} we place Gl 411 b in the context of the planets detected to date around main sequence stars within 5 pc of the Solar System. It can be seen that Gl 411 b is the second closest planet with a temperate equilibrium temperature, after Proxima Centauri b \citep{angladaescude2016a}, but it is too close to its star to have liquid water on its surface \citep{kopparapu2013}.

The maximum angular distance between the planet and its star, $a[\mathrm{AU}]/d[\mathrm{pc}]$ \citep{lovis2017}, is around 0.031 arc seconds, not unlike Proxima b, which appears at a distance of 0.037 arc seconds \citep{lovis2017}. The planet is therefore not easily resolved by the adaptive optics system on a 10-m class telescope. However, 0.031 arc seconds corresponds to approximately $6\,\lambda / D$ at $\lambda=750$ nm, on a 30-m telescope. Additionally, the flux contrast in the visible, estimated as in \citet{lovis2017} is $\left(1.04 \pm 0.08\right)\times10^{-7}$, at orbital quadrature, for a geometric albedo of 0.4 and an estimate planetary radius of 1.66 $R_\oplus$ (see above). Therefore, the technique of combining high-dispersion spectroscopy and high contrast imaging proposed originally by \citet{sparksford2002} and demonstrated by, for example, \citet{snellen2014}, \citet{schwarz2016}, and \citet{birkby2017} should be able to probe the atmosphere of this planet when the next-generation 30-m class telescopes become operational.

\begin{figure}
\input{figures/Gl411_teq_context.pgf}
\caption{Equilibrium temperature and distance to the Sun for planets currently known within 5 pc from the Solar System. Equilibrium temperatures were computed assuming circular orbits for all objects, and a Bond albedo of 0.3. Data were obtained from the exoplanets.eu \citep{schneider2011} website whenever available. The parameters of the planets are reported in \citet{angladaescude2016a, ribas2018, hatzes2000, benedict2006, bonfils2018, howard2014, feng2018, astudillodefru2017c, feng2017, astudillodefru2017b, angladaescude2014, burt2014, correia2010, rivera2010b, jenkins2014, wittenmyer2014}. The limits of the habitable zone for a star like Gl411 are indicated by green dotted lines. The symbol sizes are proportional to the natural logarithm of the minimum mass.}
\label{fig.teqcontext}
\end{figure}

\subsection{Previous planet claims. \label{sect.hires}}

The SOPHIE radial velocities presented here are clearly incompatible with the claims of companions by \citet{vandeKamp1951} and \citet{lippincott1960}, which would produce radial velocity variations of at least 1 \kms\ and of 265 \ms, respectively. Concerning the companion reported by \citet{gatewood1996}, no information is given on the inclination of the orbit, but assuming the inclination to be $i=45^\circ$ would produce a radial velocity variation of 18.5 \ms\ over the orbital period of 5.8 years. No such variation seems possible from the inspection of the RV time series in Fig.~\ref{fig.modelintime}.

On the other hand, \citet{butler2017} reported a planet candidate around Gl411 with a period of 9.8693 days and a mass of 3.8 \ME, based on long-term monitoring with the HIRES instrument. This signal is absent in the SOPHIE time series, in spite of having sufficient precision to detect it. 

\begin{figure}
\center
\input{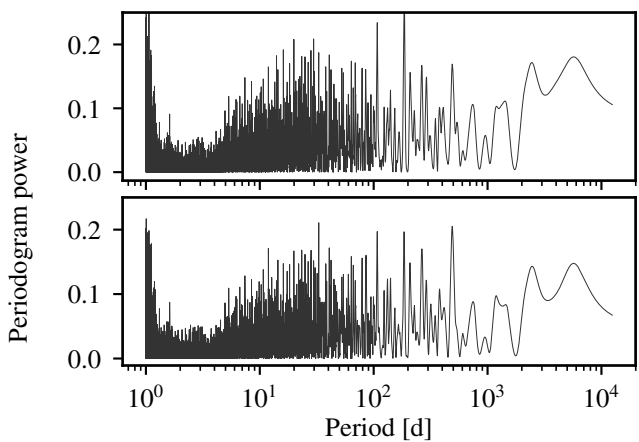}
\caption{GLS periodograms for HIRES data. \emph{Top panel:} raw HIRES data obtained from the Vizier catalogue access tool; \emph{bottom panel:} data corrected by \citet{tal-or2019}.}
\label{fig.glshires}
\end{figure}

We recovered the HIRES radial velocities available from the Vizier catalogue access tool, in order to further explore the missing signal. These are 256 measurements obtained over 17 years. After removal of a probable outlier velocity acquired on JD=2\,456\,494, the measurements have a mean uncertainty of 1.37 \ms, and a dispersion of 3.41 \ms. We also explored the HIRES data corrected by \citet{tal-or2019}, who account for a jump due to the instrument upgrade in 2004 and a slow zero-point variation. There are five additional points in this set, at the end of the series, which we disregarded for the analyses. The  corrected velocities scatter around the mean with a standard deviation of 3.35 \ms. The raw periodograms of the HIRES data (Fig.~\ref{fig.glshires}) do not show any clear periodicity at the reported period. There exists a forest of peaks between around 5 and 100 days, particularly present in the Vizier version of the data, but no clear predominant variability is observed, not the one reported by \citet{butler2017} nor the one reported here. With the corrected version of the data, the peak forest is reduced, as well as the peaks at 108 d and 185 d. A peak is seen above the noise level at around 34 days. 

The absence of the signal at 9.9-day period could be explained by the different model used by the authors, who decorrelated from the activity index based on the Ca II lines, and used a moving average model to reveal the planet. However, no predominant peak is observed in the HIRES residuals of the k0d1 model with correlated noise, described in Sect.~\ref{sect.model}, which should somehow take into account correlated variability as the one produced by activity (the resulting periodograms are very similar to those in Fig.~\ref{fig.glshires}). 

Nevertheless, we decided to perform an independent analysis in the same line as done for the SOPHIE dataset. Two models with a linear acceleration term and a Keplerian component (k1d1) were explored, one in which the {\tt emcee} walkers are started in a small region around the period reported by \citet{butler2017} --although the rest of the parameters were chosen randomly from the prior--, and a second one started close to the period detected in the SOPHIE data. We included correlated noise through a squared-exponential kernel and included an additional white-noise term, exactly as done in our analysis above. Irrespective of the data version used, in the first case, the walkers remain around the initial period of about 9.867 days, and the MAP-estimate covariance amplitude and timescale are around 2.65 \ms, and 2.03 days, respectively, which hint to additional variability in the data not explained by the Keplerian component. Under the model with a Keplerian variability initially set around the period found in SOPHIE data, the walkers expand in period space to occupy all prior space. In other words, no clear preference for the SOPHIE period is found in the HIRES data. 

It is puzzling that no sign of the signal reported here is seen in the HIRES data. However, no support is provided by the HIRES data for the other period either, independently of the version of the data considered. When the marginal likelihood is computed\footnote{Exactly the same procedure employed in the analysis of the SOPHIE data and described in Sect.~\ref{sect.analysis} was followed here.} for the model without a Keplerian curve and for the k1d1 model containing a companion at the period reported by \citet{butler2017}, we find that the model without a Keplerian curve exhibits the largest marginal likelihood. If equal prior probability is assumed for the models, this means that, under the current noise model, the model without a Keplerian signal has the largest posterior probability. The marginal likelihoods are reported in Table~\ref{table.evidencehires}. We decided not to compute the evidence for the model with the period reported in this article, because the corresponding MCMC algorithm, on whose results the computation depends, has clearly not converged.

\begin{table*}
\center
\caption{Log Marginal likelihood ($\evid$) for different models, and Bayes Factor with respect to model k0d1 for HIRES data.} \label{table.evidencehires}
\begin{tabular}{l c c c c c}        
\hline\hline                 
& & \multicolumn{2}{c}{Vizier Data} & \multicolumn{2}{c}{\citet{tal-or2019} data}\\
Model & Period [d]& $\log_{10}{\evid}$ & $\log_{10}{\mathrm{BF}_{\mathrm{k0d1}, \mathbf{\cdot}}}$ & $\log_{10}{\evid}$ & $\log_{10}{\mathrm{BF}_{\mathrm{k0d1}, \mathbf{\cdot}}}$ \\
\hline
k0d1 & -- &$-251.277 \pm 0.012$ & 0.0 				 & $-253.0089 \pm 0.0096$& 0.0\\
k1d1 & 9.867  &$-252.166\pm0.043$ & $0.889 \pm 0.045$ & $-253.193\pm0.033$	& $0.184\pm0.034$\\
\hline
\end{tabular}
\end{table*}

Of course, under a different noise model, conclusions may change. As mentioned above, the analysis of \citet{butler2017} differs from ours significantly. A detailed study of the discrepancy found and the missing signal in the SOPHIE data is deferred to a future work. We simply note that the window function of the HIRES data contains a peak exactly at the frequency of the putative planet reported by \citet{butler2017}. Although far from being the most important peak in the window function, the result by \citet{rajpaul2015} on Alpha Centauri B should cast doubts on the nature of the reported signal, specially when the velocities are known to contain low-frequency terms, such as the secular acceleration, which is here subtracted from the data before performing the analysis.

\section{Conclusion \label{sect.conclusion}}
Long-term monitoring with the SOPHIE spectrograph revealed a 2.99-Earth-mass planet around the nearby M-dwarf Gl411 (Lalande 21185, HD 95735) on a 12.95-day period orbit. Gl411 was one of the standard stars used to follow instrumental velocity systematics before it exhibited a periodic velocity variation. The signal is significantly detected in the SOPHIE data, and ancillary observations of the line bisector and width, and activity proxies failed to reveal any variability at or around the period of the putative planet. Besides, a series of simulations, assuming both correlated and uncorrelated noise show that the observed signal cannot be produced by a combination of sampling and imperfect consideration of slowly-varying terms in the model. These pieces of evidence point towards the planetary nature of the detected signal. 

The alias period at around 1.08 days is also significant, and the current data set is not fully conclusive on which is the real periodicity. However, the measured occurrence rate of very-short-period planets around low-mass stars is much lower than that of Super-earth planets on 10-day period orbits \citep{sanchisojeda2014, bonfils2013}, which further support our conclusion that the real period of the detected signal is $\sim~12.9$ days. Nevertheless, additional high-cadence observations would be useful to distinguish between both periods more precisely. 

A key component in the validation of the planetary nature of the detected signal is knowledge of the stellar rotational period. Ground-based photometric observations revealed the rotational period of Gl411 to be around 56 days. Simulations using correlated noise show that the rotational frequency is too different from that of the planetary signal to affect its detection. The planetary signal cannot be reproduced by quasi-periodic variability at the rotational frequency.

Gl411 b orbits its M2 host star at a distance of approximately 0.079 AU, which implies equilibrium temperatures between 256 K and 350 K. The planet orbits inside the inner limit of the habitable zone, and therefore seems too hot to support liquid water on its surface.

The distance to Gl411 is about 2.55 pc, making the planet Gl 411 b the third closest confirmed exoplanet. The proximity of the system, and its brightness make it an ideal target for future follow-up studies, such as searching for transits of the planet from space. Interestingly, studying the atmosphere of Gl411 b by combining high-resolution spectroscopy and high-contrast imaging, seems to be in reach of future or even current instrumentation. With the construction of 30-m class telescopes in the northern hemisphere, Gl411 b will become one of the best-suited targets for atmospheric characterisation.

\begin{acknowledgements}
We thank all the staff of Haute-Provence Observatory for their support at the 1.93-m telescope and on SOPHIE. 

We are grateful to J.F. Albacete Colombo for his help with the XMM-Newton observations of Gl411.

This work has been supported by a grant from Labex OSUG@2020 (Investissements d'avenir - ANR10 LABX56).

This work is supported by the French National Research Agency in the framework of the Investissements d'Avenir programme (ANR-15-IDEX-02).

This work is supported by the French Programme National de Physique Stellaire (PNPS) and the Programme National de Planetologie (PNP).

N. A-D. acknowledges the support of FONDECYT project 3180063.

X.B. acknowledges funding from the European Research Council under the ERC Grant Agreement n. 337591-ExTrA.

G. W. H. acknowledges long-term support from NASA, NSF, Tennessee State University, 
the State of Tennessee through its Centers of Excellence programme.

We acknowledge the support by FCT - Funda\c{c}\~ao para a Ci\^encia e a Tecnologia through national funds and by FEDER through COMPETE2020 - Programa Operacional Competitividade e Internacionaliza\c{c}\~ao by these grants: UID/FIS/04434/2013 \& POCI-01-0145-FEDER-007672; PTDC/FIS-AST/28953/2017 \& POCI-01-0145-FEDER-028953 and PTDC/FIS-AST/32113/2017 \& POCI-01-0145-FEDER-032113.

This work has been carried out within the frame of the National Centre for Competence in Research ``PlanetS'' supported by the Swiss National Science Foundation (SNSF).
V. B. and N. H. acknowledge the financial support by the Swiss National Science Foundation (SNSF) in the frame of the National Centre for Competence in Research PlanetS.
V. B. has received funding from the European Research Council (ERC) under the European Union's Horizon 2020 research and innovation programme (project Four Aces; grant agreement No 724427). 

This research has made use of the SIMBAD database, and of the VizieR catalogue access tool, CDS, Strasbourg, France. The original description of the VizieR service was published in A\&AS 143, 23.

This work used the \texttt{python} packages  \texttt{scipy} \citep{oliphant2007, millmanaivazis2011}, \texttt{pandas} \citep{mckinney2010}, and \texttt{matplotlib} \citep{hunter2007}.
\end{acknowledgements}

\bibliographystyle{aa}
\bibliography{grandbiblio}

\begin{thebibliography}{100}
\expandafter\ifx\csname natexlab\endcsname\relax\def\natexlab#1{#1}\fi

\bibitem[{{Allard} {et~al.}(2012){Allard}, {Homeier}, \&
  {Freytag}}]{allard2012}
{Allard}, F., {Homeier}, D., \& {Freytag}, B. 2012, Royal Society of London
  Philosophical Transactions Series A, 370, 2765

\bibitem[{{Allard} {et~al.}(2013){Allard}, {Homeier}, {Freytag},
  {Schaffenberger}, {}, \& {Rajpurohit}}]{allard2013}
{Allard}, F., {Homeier}, D., {Freytag}, B., {et~al.} 2013, Memorie della
  Societa Astronomica Italiana Supplementi, 24, 128

\bibitem[{{Allart} {et~al.}(2018){Allart}, {Bourrier}, {Lovis}, {Ehrenreich},
  {Spake}, {Wyttenbach}, {Pino}, {Pepe}, {Sing}, \& {Lecavelier des
  Etangs}}]{allart2018}
{Allart}, R., {Bourrier}, V., {Lovis}, C., {et~al.} 2018, arXiv e-prints
  [\eprint{1812.02189}]

\bibitem[{{Anglada-Escud{\'e}} {et~al.}(2016){Anglada-Escud{\'e}}, {Amado},
  {Barnes}, {Berdi{\~n}as}, {Butler}, {Coleman}, {de La Cueva}, {Dreizler},
  {Endl}, {Giesers}, {Jeffers}, {Jenkins}, {Jones}, {Kiraga}, {K{\"u}rster},
  {L{\'o}pez-Gonz{\'a}lez}, {Marvin}, {Morales}, {Morin}, {Nelson}, {Ortiz},
  {Ofir}, {Paardekooper}, {Reiners}, {Rodr{\'{\i}}guez},
  {Rodr{\'{\i}}guez-L{\'o}pez}, {Sarmiento}, {Strachan}, {Tsapras}, {Tuomi}, \&
  {Zechmeister}}]{angladaescude2016a}
{Anglada-Escud{\'e}}, G., {Amado}, P.~J., {Barnes}, J., {et~al.} 2016, \nat,
  536, 437

\bibitem[{{Anglada-Escud{\'e}} {et~al.}(2014){Anglada-Escud{\'e}}, {Arriagada},
  {Tuomi}, {Zechmeister}, {Jenkins}, {Ofir}, {Dreizler}, {Gerlach}, {Marvin},
  {Reiners}, {Jeffers}, {Butler}, {Vogt}, {Amado},
  {Rodr{\'{\i}}guez-L{\'o}pez}, {Berdi{\~n}as}, {Morin}, {Crane}, {Shectman},
  {Thompson}, {D{\'{\i}}az}, {Rivera}, {Sarmiento}, \&
  {Jones}}]{angladaescude2014}
{Anglada-Escud{\'e}}, G., {Arriagada}, P., {Tuomi}, M., {et~al.} 2014, \mnras,
  443, L89

\bibitem[{{Astudillo-Defru} {et~al.}(2015){Astudillo-Defru}, {Bonfils},
  {Delfosse}, {S{\'e}gransan}, {Forveille}, {Bouchy}, {Gillon}, {Lovis},
  {Mayor}, {Neves}, {Pepe}, {Perrier}, {Queloz}, {Rojo}, {Santos}, \&
  {Udry}}]{astudillodefru2015}
{Astudillo-Defru}, N., {Bonfils}, X., {Delfosse}, X., {et~al.} 2015, \aap, 575,
  A119

\bibitem[{{Astudillo-Defru} {et~al.}(2017{\natexlab{a}}){Astudillo-Defru},
  {Delfosse}, {Bonfils}, {Forveille}, {Lovis}, \&
  {Rameau}}]{astudillodefru2017a}
{Astudillo-Defru}, N., {Delfosse}, X., {Bonfils}, X., {et~al.}
  2017{\natexlab{a}}, \aap, 600, A13

\bibitem[{{Astudillo-Defru} {et~al.}(2017{\natexlab{b}}){Astudillo-Defru},
  {D{\'\i}az}, {Bonfils}, {Almenara}, {Delisle}, {Bouchy}, {Delfosse},
  {Forveille}, {Lovis}, {Mayor}, {Murgas}, {Pepe}, {Santos}, {S{\'e}gransan},
  {Udry}, \& {W{\"u}nsche}}]{astudillodefru2017c}
{Astudillo-Defru}, N., {D{\'\i}az}, R.~F., {Bonfils}, X., {et~al.}
  2017{\natexlab{b}}, \aap, 605, L11

\bibitem[{{Astudillo-Defru} {et~al.}(2017{\natexlab{c}}){Astudillo-Defru},
  {Forveille}, {Bonfils}, {S{\'e}gransan}, {Bouchy}, {Delfosse}, {Lovis},
  {Mayor}, {Murgas}, {Pepe}, {Santos}, {Udry}, \&
  {W{\"u}nsche}}]{astudillodefru2017b}
{Astudillo-Defru}, N., {Forveille}, T., {Bonfils}, X., {et~al.}
  2017{\natexlab{c}}, \aap, 602, A88

\bibitem[{{Beichman} {et~al.}(2014){Beichman}, {Benneke}, {Knutson}, {Smith},
  {Lagage}, {Dressing}, {Latham}, {Lunine}, {Birkmann}, {Ferruit}, {Giardino},
  {Kempton}, {Carey}, {Krick}, {Deroo}, {Mandell}, {Ressler}, {Shporer},
  {Swain}, {Vasisht}, {Ricker}, {Bouwman}, {Crossfield}, {Greene}, {Howell},
  {Christiansen}, {Ciardi}, {Clampin}, {Greenhouse}, {Sozzetti}, {Goudfrooij},
  {Hines}, {Keyes}, {Lee}, {McCullough}, {Robberto}, {Stansberry}, {Valenti},
  {Rieke}, {Rieke}, {Fortney}, {Bean}, {Kreidberg}, {Ehrenreich}, {Deming},
  {Albert}, {Doyon}, \& {Sing}}]{beichman2014}
{Beichman}, C., {Benneke}, B., {Knutson}, H., {et~al.} 2014, \pasp, 126, 1134

\bibitem[{{Benedict} {et~al.}(2006){Benedict}, {McArthur}, {Gatewood}, {Nelan},
  {Cochran}, {Hatzes}, {Endl}, {Wittenmyer}, {Baliunas}, {Walker}, {Yang},
  {K{\"u}rster}, {Els}, \& {Paulson}}]{benedict2006}
{Benedict}, G.~F., {McArthur}, B.~E., {Gatewood}, G., {et~al.} 2006, \aj, 132,
  2206

\bibitem[{{Birkby} {et~al.}(2017){Birkby}, {de Kok}, {Brogi}, {Schwarz}, \&
  {Snellen}}]{birkby2017}
{Birkby}, J.~L., {de Kok}, R.~J., {Brogi}, M., {Schwarz}, H., \& {Snellen},
  I.~A.~G. 2017, \aj, 153, 138

\bibitem[{{Boisse} {et~al.}(2011){Boisse}, {Bouchy}, {Chazelas}, {Perruchot},
  {Pepe}, {Lovis}, \& {H{\'e}brard}}]{boisse2011b}
{Boisse}, I., {Bouchy}, F., {Chazelas}, B., {et~al.} 2011, Research, Science
  and Technology of Brown Dwarfs and Exoplanets: Proceedings of an
  International Conference held in Shangai on Occasion of a Total Eclipse of
  the Sun, Shangai, China, Edited by E.L.~Martin; J.~Ge; W.~Lin; EPJ Web of
  Conferences, Volume 16, id.02003, 16, 2003

\bibitem[{{Boisse} {et~al.}(2010){Boisse}, {Eggenberger}, {Santos}, {Lovis},
  {Bouchy}, {H{\'e}brard}, {Arnold}, {Bonfils}, {Delfosse}, {Desort},
  {D{\'{\i}}az}, {Ehrenreich}, {Forveille}, {Gallenne}, {Lagrange}, {Moutou},
  {Udry}, {Pepe}, {Perrier}, {Perruchot}, {Pont}, {Queloz}, {Santerne},
  {S{\'e}gransan}, \& {Vidal-Madjar}}]{boisse2010}
{Boisse}, I., {Eggenberger}, A., {Santos}, N.~C., {et~al.} 2010, \aap, 523,
  A88+

\bibitem[{{Boisse} {et~al.}(2009){Boisse}, {Moutou}, {Vidal-Madjar}, {Bouchy},
  {Pont}, {H{\'e}brard}, {Bonfils}, {Croll}, {Delfosse}, {Desort}, {Forveille},
  {Lagrange}, {Loeillet}, {Lovis}, {Matthews}, {Mayor}, {Pepe}, {Perrier},
  {Queloz}, {Rowe}, {Santos}, {S{\'e}gransan}, \& {Udry}}]{boisse2009}
{Boisse}, I., {Moutou}, C., {Vidal-Madjar}, A., {et~al.} 2009, \aap, 495, 959

\bibitem[{Bonfils {et~al.}(2018)Bonfils, Astudillo-Defru, D{\'{\i}}az,
  Almenara, Forveille, Bouchy, Delfosse, Lovis, Mayor, Murgas, Pepe, Santos,
  S{\'e}gransan, Udry, \& W{\"u}nsche}]{bonfils2018}
Bonfils, X., Astudillo-Defru, N., D{\'{\i}}az, R., {et~al.} 2018, \aap, 613,
  A25

\bibitem[{{Bonfils} {et~al.}(2013){Bonfils}, {Delfosse}, {Udry}, {Forveille},
  {Mayor}, {Perrier}, {Bouchy}, {Gillon}, {Lovis}, {Pepe}, {Queloz}, {Santos},
  {S{\'e}gransan}, \& {Bertaux}}]{bonfils2013}
{Bonfils}, X., {Delfosse}, X., {Udry}, S., {et~al.} 2013, \aap, 549, A109

\bibitem[{{Bouchy} {et~al.}(2013){Bouchy}, {D{\'{\i}}az}, {H{\'e}brard},
  {Arnold}, {Boisse}, {Delfosse}, {Perruchot}, \& {Santerne}}]{bouchy2013}
{Bouchy}, F., {D{\'{\i}}az}, R.~F., {H{\'e}brard}, G., {et~al.} 2013, \aap,
  549, A49

\bibitem[{{Bouchy} {et~al.}(2009{\natexlab{a}}){Bouchy}, {H{\'e}brard}, {Udry},
  {Delfosse}, {Boisse}, {Desort}, {Bonfils}, {Eggenberger}, {Ehrenreich},
  {Forveille}, {Lagrange}, {Le Coroller}, {Lovis}, {Moutou}, {Pepe}, {Perrier},
  {Pont}, {Queloz}, {Santos}, {S{\'e}gransan}, \& {Vidal-Madjar}}]{bouchy2009}
{Bouchy}, F., {H{\'e}brard}, G., {Udry}, S., {et~al.} 2009{\natexlab{a}}, \aap,
  505, 853

\bibitem[{{Bouchy} {et~al.}(2009{\natexlab{b}}){Bouchy}, {Isambert}, {Lovis},
  {Boisse}, {Figueira}, {H{\'e}brard}, \& {Pepe}}]{bouchy2009c}
{Bouchy}, F., {Isambert}, J., {Lovis}, C., {et~al.} 2009{\natexlab{b}}, in EAS
  Publications Series, Vol.~37, EAS Publications Series, ed. P.~{Kern},
  247--253

\bibitem[{{Broeg} {et~al.}(2013){Broeg}, {Fortier}, {Ehrenreich}, {Alibert},
  {Baumjohann}, {Benz}, {Deleuil}, {Gillon}, {Ivanov}, {Liseau}, {Meyer},
  {Oloffson}, {Pagano}, {Piotto}, {Pollacco}, {Queloz}, {Ragazzoni}, {Renotte},
  {Steller}, \& {Thomas}}]{broeg2013}
{Broeg}, C., {Fortier}, A., {Ehrenreich}, D., {et~al.} 2013, in European
  Physical Journal Web of Conferences, Vol.~47, European Physical Journal Web
  of Conferences, 03005

\bibitem[{{Burt} {et~al.}(2014){Burt}, {Vogt}, {Butler}, {Hanson}, {Meschiari},
  {Rivera}, {Henry}, \& {Laughlin}}]{burt2014}
{Burt}, J., {Vogt}, S.~S., {Butler}, R.~P., {et~al.} 2014, \apj, 789, 114

\bibitem[{{Butler} {et~al.}(2017){Butler}, {Vogt}, {Laughlin}, {Burt},
  {Rivera}, {Tuomi}, {Teske}, {Arriagada}, {Diaz}, {Holden}, \&
  {Keiser}}]{butler2017}
{Butler}, R.~P., {Vogt}, S.~S., {Laughlin}, G., {et~al.} 2017, \aj, 153, 208

\bibitem[{{Butler} {et~al.}(2004){Butler}, {Vogt}, {Marcy}, {Fischer},
  {Wright}, {Henry}, {Laughlin}, \& {Lissauer}}]{butler2004}
{Butler}, R.~P., {Vogt}, S.~S., {Marcy}, G.~W., {et~al.} 2004, \apj, 617, 580

\bibitem[{{Chen} \& {Kipping}(2017)}]{chenkipping2017}
{Chen}, J. \& {Kipping}, D. 2017, \apj, 834, 17

\bibitem[{Correia {et~al.}(2010)Correia, Couetdic, Laskar, Bonfils, Mayor,
  Bertaux, Bouchy, Delfosse, Forveille, Lovis, Pepe, Perrier, Queloz, \&
  Udry}]{correia2010}
Correia, A. C.~M., Couetdic, J., Laskar, J., {et~al.} 2010, \aap, 511, A21

\bibitem[{{Courcol} {et~al.}(2015){Courcol}, {Bouchy}, {Pepe}, {Santerne},
  {Delfosse}, {Arnold}, {Astudillo-Defru}, {Boisse}, {Bonfils}, {Borgniet},
  {Bourrier}, {Cabrera}, {Deleuil}, {Demangeon}, {D{\'{\i}}az}, {Ehrenreich},
  {Forveille}, {H{\'e}brard}, {Lagrange}, {Montagnier}, {Moutou}, {Rey},
  {Santos}, {S{\'e}gransan}, {Udry}, \& {Wilson}}]{courcol2015}
{Courcol}, B., {Bouchy}, F., {Pepe}, F., {et~al.} 2015, \aap, 581, A38

\bibitem[{{Cutri} {et~al.}(2003){Cutri}, {Skrutskie}, {van Dyk}, {Beichman},
  {Carpenter}, {Chester}, {Cambresy}, {Evans}, {Fowler}, {Gizis}, {Howard},
  {Huchra}, {Jarrett}, {Kopan}, {Kirkpatrick}, {Light}, {Marsh}, {McCallon},
  {Schneider}, {Stiening}, {Sykes}, {Weinberg}, {Wheaton}, {Wheelock}, \&
  {Zacarias}}]{2mass}
{Cutri}, R.~M., {Skrutskie}, M.~F., {van Dyk}, S., {et~al.} 2003, {2MASS All
  Sky Catalog of point sources.}

\bibitem[{{Dawson} \& {Fabrycky}(2010)}]{dawsonfabrycky2010}
{Dawson}, R.~I. \& {Fabrycky}, D.~C. 2010, \apj, 722, 937

\bibitem[{{Delfosse} {et~al.}(1998){Delfosse}, {Forveille}, {Perrier}, \&
  {Mayor}}]{delfosse98}
{Delfosse}, X., {Forveille}, T., {Perrier}, C., \& {Mayor}, M. 1998, \aap, 331,
  581

\bibitem[{{Delfosse} {et~al.}(2000){Delfosse}, {Forveille}, {S{\'e}gransan},
  {Beuzit}, {Udry}, {Perrier}, \& {Mayor}}]{delfosse2000}
{Delfosse}, X., {Forveille}, T., {S{\'e}gransan}, D., {et~al.} 2000, \aap, 364,
  217

\bibitem[{{D{\'{\i}}az} {et~al.}(2012){D{\'{\i}}az}, {Santerne}, {Sahlmann},
  {H{\'e}brard}, {Eggenberger}, {Santos}, {Moutou}, {Arnold}, {Boisse},
  {Bonfils}, {Bouchy}, {Delfosse}, {Desort}, {Ehrenreich}, {Forveille},
  {Lagrange}, {Lovis}, {Pepe}, {Perrier}, {Queloz}, {S{\'e}gransan}, {Udry}, \&
  {Vidal-Madjar}}]{diaz2012}
{D{\'{\i}}az}, R.~F., {Santerne}, A., {Sahlmann}, J., {et~al.} 2012, \aap, 538,
  A113

\bibitem[{{D{\'{\i}}az} {et~al.}(2016){D{\'{\i}}az}, {S{\'e}gransan}, {Udry},
  {Lovis}, {Pepe}, {Dumusque}, {Marmier}, {Alonso}, {Benz}, {Bouchy},
  {Coffinet}, {Collier Cameron}, {Deleuil}, {Figueira}, {Gillon}, {Lo Curto},
  {Mayor}, {Mordasini}, {Motalebi}, {Moutou}, {Pollacco}, {Pompei}, {Queloz},
  {Santos}, \& {Wyttenbach}}]{diaz2016a}
{D{\'{\i}}az}, R.~F., {S{\'e}gransan}, D., {Udry}, S., {et~al.} 2016, \aap,
  585, A134

\bibitem[{{Doyon} {et~al.}(2014){Doyon}, {Lafreni{\`e}re}, {Albert}, {Artigau},
  {Meyer}, \& {Jayawardhana}}]{doyon2014}
{Doyon}, R., {Lafreni{\`e}re}, D., {Albert}, L., {et~al.} 2014, in Search for
  Life Beyond the Solar System. Exoplanets, Biosignatures \& Instruments, ed.
  D.~{Apai} \& P.~{Gabor}, 3.6

\bibitem[{{Dressing} \& {Charbonneau}(2015)}]{dressingcharbonneau2015}
{Dressing}, C.~D. \& {Charbonneau}, D. 2015, \apj, 807, 45

\bibitem[{{Eaton} {et~al.}(2003){Eaton}, {Henry}, \& {Fekel}}]{eaton2003}
{Eaton}, J.~A., {Henry}, G.~W., \& {Fekel}, F.~C. 2003, in Astrophysics and
  Space Science Library, Vol. 288, Astrophysics and Space Science Library, ed.
  T.~D. {Oswalt}, 189

\bibitem[{{F.~van Leeuwen}(2007)}]{vanleeuwen2007}
{F.~van Leeuwen}, ed. 2007, Astrophysics and Space Science Library, Vol. 350,
  {Hipparcos, the New Reduction of the Raw Data}

\bibitem[{{Feng} {et~al.}(2018){Feng}, {Tuomi}, \& {Jones}}]{feng2018}
{Feng}, F., {Tuomi}, M., \& {Jones}, H.~R.~A. 2018, arXiv e-prints
  [\eprint[arXiv]{1803.08163}]

\bibitem[{{Feng} {et~al.}(2017){Feng}, {Tuomi}, {Jones}, {Barnes},
  {Anglada-Escud{\'e}}, {Vogt}, \& {Butler}}]{feng2017}
{Feng}, F., {Tuomi}, M., {Jones}, H.~R.~A., {et~al.} 2017, \aj, 154, 135

\bibitem[{{Foreman-Mackey} {et~al.}(2013){Foreman-Mackey}, {Hogg}, {Lang}, \&
  {Goodman}}]{emcee}
{Foreman-Mackey}, D., {Hogg}, D.~W., {Lang}, D., \& {Goodman}, J. 2013, \pasp,
  125, 306

\bibitem[{{Gaia Collaboration} {et~al.}(2018){Gaia Collaboration}, {Brown},
  {Vallenari}, {Prusti}, {de Bruijne}, {Babusiaux}, {Bailer-Jones}, {Biermann},
  {Evans}, {Eyer}, {Jansen}, {Jordi}, {Klioner}, {Lammers}, {Lindegren},
  {Luri}, {Mignard}, {Panem}, {Pourbaix}, {Randich}, {Sartoretti}, {Siddiqui},
  {Soubiran}, {van Leeuwen}, {Walton}, {Arenou}, {Bastian}, {Cropper},
  {Drimmel}, {Katz}, {Lattanzi}, {Bakker}, {Cacciari}, {Casta{\~n}eda},
  {Chaoul}, {Cheek}, {De Angeli}, {Fabricius}, {Guerra}, {Holl}, {Masana},
  {Messineo}, {Mowlavi}, {Nienartowicz}, {Panuzzo}, {Portell}, {Riello},
  {Seabroke}, {Tanga}, {Th{\'e}venin}, {Gracia-Abril}, {Comoretto},
  {Garcia-Reinaldos}, {Teyssier}, {Altmann}, {Andrae}, {Audard},
  {Bellas-Velidis}, {Benson}, {Berthier}, {Blomme}, {Burgess}, {Busso},
  {Carry}, {Cellino}, {Clementini}, {Clotet}, {Creevey}, {Davidson}, {De
  Ridder}, {Delchambre}, {Dell'Oro}, {Ducourant}, {Fern{\'a}ndez-
  Hern{\'a}ndez}, {Fouesneau}, {Fr{\'e}mat}, {Galluccio}, {Garc{\'\i}a-Torres},
  {Gonz{\'a}lez-N{\'u}{\~n}ez}, {Gonz{\'a}lez-Vidal}, {Gosset}, {Guy},
  {Halbwachs}, {Hambly}, {Harrison}, {Hern{\'a}ndez}, {Hestroffer}, {Hodgkin},
  {Hutton}, {Jasniewicz}, {Jean-Antoine-Piccolo}, {Jordan}, {Korn},
  {Krone-Martins}, {Lanzafame}, {Lebzelter}, {L{\"o}ffler}, {Manteiga},
  {Marrese}, {Mart{\'\i}n-Fleitas}, {Moitinho}, {Mora}, {Muinonen}, {Osinde},
  {Pancino}, {Pauwels}, {Petit}, {Recio-Blanco}, {Richards}, {Rimoldini},
  {Robin}, {Sarro}, {Siopis}, {Smith}, {Sozzetti}, {S{\"u}veges}, {Torra}, {van
  Reeven}, {Abbas}, {Abreu Aramburu}, {Accart}, {Aerts}, {Altavilla},
  {{\'A}lvarez}, {Alvarez}, {Alves}, {Anderson}, {Andrei}, {Anglada Varela},
  {Antiche}, {Antoja}, {Arcay}, {Astraatmadja}, {Bach}, {Baker},
  {Balaguer-N{\'u}{\~n}ez}, {Balm}, {Barache}, {Barata}, {Barbato}, {Barblan},
  {Barklem}, {Barrado}, {Barros}, {Barstow}, {Bartholom{\'e} Mu{\~n}oz},
  {Bassilana}, {Becciani}, {Bellazzini}, {Berihuete}, {Bertone}, {Bianchi},
  {Bienaym{\'e}}, {Blanco-Cuaresma}, {Boch}, {Boeche}, {Bombrun}, {Borrachero},
  {Bossini}, {Bouquillon}, {Bourda}, {Bragaglia}, {Bramante}, {Breddels},
  {Bressan}, {Brouillet}, {Br{\"u}semeister}, {Brugaletta}, {Bucciarelli},
  {Burlacu}, {Busonero}, {Butkevich}, {Buzzi}, {Caffau}, {Cancelliere},
  {Cannizzaro}, {Cantat-Gaudin}, {Carballo}, {Carlucci}, {Carrasco},
  {Casamiquela}, {Castellani}, {Castro-Ginard}, {Charlot}, {Chemin},
  {Chiavassa}, {Cocozza}, {Costigan}, {Cowell}, {Crifo}, {Crosta}, {Crowley},
  {Cuypers}, {Dafonte}, {Damerdji}, {Dapergolas}, {David}, {David}, {de
  Laverny}, {De Luise}, {De March}, {de Martino}, {de Souza}, {de Torres},
  {Debosscher}, {del Pozo}, {Delbo}, {Delgado}, {Delgado}, {Di Matteo},
  {Diakite}, {Diener}, {Distefano}, {Dolding}, {Drazinos}, {Dur{\'a}n},
  {Edvardsson}, {Enke}, {Eriksson}, {Esquej}, {Eynard Bontemps}, {Fabre},
  {Fabrizio}, {Faigler}, {Falc{\~a}o}, {Farr{\`a}s Casas}, {Federici},
  {Fedorets}, {Fernique}, {Figueras}, {Filippi}, {Findeisen}, {Fonti},
  {Fraile}, {Fraser}, {Fr{\'e}zouls}, {Gai}, {Galleti}, {Garabato},
  {Garc{\'\i}a-Sedano}, {Garofalo}, {Garralda}, {Gavel}, {Gavras}, {Gerssen},
  {Geyer}, {Giacobbe}, {Gilmore}, {Girona}, {Giuffrida}, {Glass}, {Gomes},
  {Granvik}, {Gueguen}, {Guerrier}, {Guiraud}, {Guti{\'e}rrez-S{\'a}nchez},
  {Haigron}, {Hatzidimitriou}, {Hauser}, {Haywood}, {Heiter}, {Helmi}, {Heu},
  {Hilger}, {Hobbs}, {Hofmann}, {Holland}, {Huckle}, {Hypki}, {Icardi},
  {Jan{\ss}en}, {Jevardat de Fombelle}, {Jonker}, {Juh{\'a}sz}, {Julbe},
  {Karampelas}, {Kewley}, {Klar}, {Kochoska}, {Kohley}, {Kolenberg},
  {Kontizas}, {Kontizas}, {Koposov}, {Kordopatis}, {Kostrzewa-Rutkowska},
  {Koubsky}, {Lambert}, {Lanza}, {Lasne}, {Lavigne}, {Le Fustec}, {Le
  Poncin-Lafitte}, {Lebreton}, {Leccia}, {Leclerc}, {Lecoeur-Taibi},
  {Lenhardt}, {Leroux}, {Liao}, {Licata}, {Lindstr{\o}m}, {Lister}, {Livanou},
  {Lobel}, {L{\'o}pez}, {Managau}, {Mann}, {Mantelet}, {Marchal}, {Marchant},
  {Marconi}, {Marinoni}, {Marschalk{\'o}}, {Marshall}, {Martino}, {Marton},
  {Mary}, {Massari}, {Matijevi{\v{c}}}, {Mazeh}, {McMillan}, {Messina},
  {Michalik}, {Millar}, {Molina}, {Molinaro}, {Moln{\'a}r}, {Montegriffo},
  {Mor}, {Morbidelli}, {Morel}, {Morris}, {Mulone}, {Muraveva}, {Musella},
  {Nelemans}, {Nicastro}, {Noval}, {O'Mullane}, {Ord{\'e}novic},
  {Ord{\'o}{\~n}ez-Blanco}, {Osborne}, {Pagani}, {Pagano}, {Pailler},
  {Palacin}, {Palaversa}, {Panahi}, {Pawlak}, {Piersimoni}, {Pineau}, {Plachy},
  {Plum}, {Poggio}, {Poujoulet}, {Pr{\v{s}}a}, {Pulone}, {Racero}, {Ragaini},
  {Rambaux}, {Ramos-Lerate}, {Regibo}, {Reyl{\'e}}, {Riclet}, {Ripepi}, {Riva},
  {Rivard}, {Rixon}, {Roegiers}, {Roelens}, {Romero-G{\'o}mez}, {Rowell},
  {Royer}, {Ruiz-Dern}, {Sadowski}, {Sagrist{\`a} Sell{\'e}s}, {Sahlmann},
  {Salgado}, {Salguero}, {Sanna}, {Santana- Ros}, {Sarasso}, {Savietto},
  {Schultheis}, {Sciacca}, {Segol}, {Segovia}, {S{\'e}gransan}, {Shih},
  {Siltala}, {Silva}, {Smart}, {Smith}, {Solano}, {Solitro}, {Sordo}, {Soria
  Nieto}, {Souchay}, {Spagna}, {Spoto}, {Stampa}, {Steele},
  {Steidelm{\"u}ller}, {Stephenson}, {Stoev}, {Suess}, {Surdej}, {Szabados},
  {Szegedi-Elek}, {Tapiador}, {Taris}, {Tauran}, {Taylor}, {Teixeira},
  {Terrett}, {Teyssandier}, {Thuillot}, {Titarenko}, {Torra Clotet}, {Turon},
  {Ulla}, {Utrilla}, {Uzzi}, {Vaillant}, {Valentini}, {Valette}, {van Elteren},
  {Van Hemelryck}, {van Leeuwen}, {Vaschetto}, {Vecchiato}, {Veljanoski},
  {Viala}, {Vicente}, {Vogt}, {von Essen}, {Voss}, {Votruba}, {Voutsinas},
  {Walmsley}, {Weiler}, {Wertz}, {Wevers}, {Wyrzykowski}, {Yoldas},
  {{\v{Z}}erjal}, {Ziaeepour}, {Zorec}, {Zschocke}, {Zucker}, {Zurbach}, \&
  {Zwitter}}]{gaiaDR2}
{Gaia Collaboration}, {Brown}, A.~G.~A., {Vallenari}, A., {et~al.} 2018, \aap,
  616, A1

\bibitem[{{Gaidos} {et~al.}(2016){Gaidos}, {Mann}, {Kraus}, \&
  {Ireland}}]{gaidos2016}
{Gaidos}, E., {Mann}, A.~W., {Kraus}, A.~L., \& {Ireland}, M. 2016, \mnras,
  457, 2877

\bibitem[{{Gaidos} {et~al.}(2014){Gaidos}, {Mann}, {L{\'e}pine}, {Buccino},
  {James}, {Ansdell}, {Petrucci}, {Mauas}, \& {Hilton}}]{gaidos2014}
{Gaidos}, E., {Mann}, A.~W., {L{\'e}pine}, S., {et~al.} 2014, \mnras, 443, 2561

\bibitem[{{Gatewood}(1996)}]{gatewood1996}
{Gatewood}, G. 1996, in American Astronomical Society Meeting Abstracts, Vol.
  188, American Astronomical Society Meeting Abstracts \#188, 40.11

\bibitem[{Goodman \& Weare(2010)}]{goodmanweare2010}
Goodman, J. \& Weare, J. 2010, Communications in applied mathematics and
  computational science, 5, 65

\bibitem[{{Handley} {et~al.}(2015){Handley}, {Hobson}, \&
  {Lasenby}}]{handley2015}
{Handley}, W.~J., {Hobson}, M.~P., \& {Lasenby}, A.~N. 2015, \mnras, 453, 4384

\bibitem[{{Hatzes} {et~al.}(2000){Hatzes}, {Cochran}, {McArthur}, {Baliunas},
  {Walker}, {Campbell}, {Irwin}, {Yang}, {K{\"u}rster}, {Endl}, {Els},
  {Butler}, \& {Marcy}}]{hatzes2000}
{Hatzes}, A.~P., {Cochran}, W.~D., {McArthur}, B., {et~al.} 2000, \apjl, 544,
  L145

\bibitem[{{Haywood} {et~al.}(2014){Haywood}, {Collier Cameron}, {Queloz},
  {Barros}, {Deleuil}, {Fares}, {Gillon}, {Lanza}, {Lovis}, {Moutou}, {Pepe},
  {Pollacco}, {Santerne}, {S{\'e}gransan}, \& {Unruh}}]{haywood2014}
{Haywood}, R.~D., {Collier Cameron}, A., {Queloz}, D., {et~al.} 2014, \mnras,
  443, 2517

\bibitem[{{Henry}(1995{\natexlab{a}})}]{henry1995b}
{Henry}, G.~W. 1995{\natexlab{a}}, in Astronomical Society of the Pacific
  Conference Series, Vol.~79, Robotic Telescopes. Current Capabilities, Present
  Developments, and Future Prospects for Automated Astronomy, ed. G.~W. {Henry}
  \& J.~A. {Eaton}, 44

\bibitem[{{Henry}(1995{\natexlab{b}})}]{henry1995a}
{Henry}, G.~W. 1995{\natexlab{b}}, in Astronomical Society of the Pacific
  Conference Series, Vol.~79, Robotic Telescopes. Current Capabilities, Present
  Developments, and Future Prospects for Automated Astronomy, ed. G.~W. {Henry}
  \& J.~A. {Eaton}, 37

\bibitem[{{Henry}(1999)}]{henry1999}
{Henry}, G.~W. 1999, \pasp, 111, 845

\bibitem[{{Hobson} {et~al.}(2018){Hobson}, {D{\'{\i}}az}, {Delfosse},
  {Astudillo-Defru}, {Boisse}, {Bouchy}, {Bonfils}, {Forveille}, {Hara},
  {Arnold}, {Borgniet}, {Bourrier}, {Brugger}, {Cabrera}, {Courcol}, {Dalal},
  {Deleuil}, {Demangeon}, {Dumusque}, {Ehrenreich}, {H{\'e}brard}, {Kiefer},
  {Lopez}, {Mignon}, {Montagnier}, {Mousis}, {Moutou}, {Pepe}, {Rey},
  {Santerne}, {Santos}, {Stalport}, {S{\'e}gransan}, {Udry}, \&
  {Wilson}}]{hobson2018}
{Hobson}, M.~J., {D{\'{\i}}az}, R.~F., {Delfosse}, X., {et~al.} 2018, \aap,
  618, A103

\bibitem[{Houdebine(2010)}]{houdebine2010}
Houdebine, E.~R. 2010, \mnras, 407, 1657

\bibitem[{{Howard} {et~al.}(2014){Howard}, {Marcy}, {Fischer}, {Isaacson},
  {Muirhead}, {Henry}, {Boyajian}, {von Braun}, {Becker}, {Wright}, \&
  {Johnson}}]{howard2014}
{Howard}, A.~W., {Marcy}, G.~W., {Fischer}, D.~A., {et~al.} 2014, \apj, 794, 51

\bibitem[{Hunter(2007)}]{hunter2007}
Hunter, J.~D. 2007, Computing in Science Engineering, 9, 90

\bibitem[{{Jansen} {et~al.}(2001){Jansen}, {Lumb}, {Altieri}, {Clavel}, {Ehle},
  {Erd}, {Gabriel}, {Guainazzi}, {Gondoin}, {Much}, {Munoz}, {Santos},
  {Schartel}, {Texier}, \& {Vacanti}}]{jansen2001}
{Jansen}, F., {Lumb}, D., {Altieri}, B., {et~al.} 2001, \aap, 365, L1

\bibitem[{{Jenkins} {et~al.}(2014){Jenkins}, {Yoma}, {Rojo}, {Mahu}, \&
  {Wuth}}]{jenkins2014}
{Jenkins}, J.~S., {Yoma}, N.~B., {Rojo}, P., {Mahu}, R., \& {Wuth}, J. 2014,
  \mnras, 441, 2253

\bibitem[{Kass \& Raftery(1995)}]{kassraftery1995}
Kass, R.~E. \& Raftery, A.~E. 1995, Journal of the american statistical
  association, 90, 773

\bibitem[{{Kiraga} \& {Stepien}(2007)}]{kiragastepien2007}
{Kiraga}, M. \& {Stepien}, K. 2007, \actaa, 57, 149

\bibitem[{{Kopparapu} {et~al.}(2013){Kopparapu}, {Ramirez}, {Kasting}, {Eymet},
  {Robinson}, {Mahadevan}, {Terrien}, {Domagal-Goldman}, {Meadows}, \&
  {Deshpande}}]{kopparapu2013}
{Kopparapu}, R.~K., {Ramirez}, R., {Kasting}, J.~F., {et~al.} 2013, \apj, 765,
  131

\bibitem[{K{\"u}rster {et~al.}(2003)K{\"u}rster, Endl, Rouesnel, Els, Kaufer,
  Brillant, Hatzes, Saar, \& Cochran}]{kurster2003}
K{\"u}rster, M., Endl, M., Rouesnel, F., {et~al.} 2003, \aap, 403, 1077

\bibitem[{{L{\'e}pine} \& {Gaidos}(2011)}]{lepinegaidos2011}
{L{\'e}pine}, S. \& {Gaidos}, E. 2011, \aj, 142, 138

\bibitem[{{Lippincott}(1960)}]{lippincott1960}
{Lippincott}, S.~L. 1960, \aj, 65, 445

\bibitem[{{Lovis} {et~al.}(2017){Lovis}, {Snellen}, {Mouillet}, {Pepe},
  {Wildi}, {Astudillo-Defru}, {Beuzit}, {Bonfils}, {Cheetham}, {Conod},
  {Delfosse}, {Ehrenreich}, {Figueira}, {Forveille}, {Martins}, {Quanz},
  {Santos}, {Schmid}, {S{\'e}gransan}, \& {Udry}}]{lovis2017}
{Lovis}, C., {Snellen}, I., {Mouillet}, D., {et~al.} 2017, \aap, 599, A16

\bibitem[{{Mann} {et~al.}(2015){Mann}, {Feiden}, {Gaidos}, {Boyajian}, \& {von
  Braun}}]{mann2015}
{Mann}, A.~W., {Feiden}, G.~A., {Gaidos}, E., {Boyajian}, T., \& {von Braun},
  K. 2015, \apj, 804, 64

\bibitem[{McKinney(2010)}]{mckinney2010}
McKinney, W. 2010, in Proceedings of the 9th Python in Science Conference, ed.
  S.~van~der Walt \& J.~Millman, 51 -- 56

\bibitem[{Millman \& Aivazis(2011)}]{millmanaivazis2011}
Millman, K.~J. \& Aivazis, M. 2011, Computing in Science Engineering, 13, 9

\bibitem[{{Mortier} \& {Collier Cameron}(2017)}]{mortiercameron2017}
{Mortier}, A. \& {Collier Cameron}, A. 2017, \aap, 601, A110

\bibitem[{{Murray} \& {Dermott}(2000)}]{murraydermott2000}
{Murray}, C.~D. \& {Dermott}, S.~F. 2000, {Solar System Dynamics} ({Cambridge
  University Press})

\bibitem[{{Nelson} {et~al.}(2018){Nelson}, {Ford}, {Buchner}, {Cloutier},
  {D{\'{\i}}az}, {Faria}, {Rajpaul}, \& {Rukdee}}]{nelson2018}
{Nelson}, B.~E., {Ford}, E.~B., {Buchner}, J., {et~al.} 2018, ArXiv e-prints
  [\eprint[\apj, submited]{1806.04683}]

\bibitem[{{Noyes} {et~al.}(1984){Noyes}, {Hartmann}, {Baliunas}, {Duncan}, \&
  {Vaughan}}]{noyes84}
{Noyes}, R.~W., {Hartmann}, L.~W., {Baliunas}, S.~L., {Duncan}, D.~K., \&
  {Vaughan}, A.~H. 1984, \apj, 279, 763

\bibitem[{Oliphant(2007)}]{oliphant2007}
Oliphant, T.~E. 2007, Computing in Science Engineering, 9, 10

\bibitem[{Perrakis {et~al.}(2014)Perrakis, Ntzoufras, \&
  Tsionas}]{perrakis2014}
Perrakis, K., Ntzoufras, I., \& Tsionas, E.~G. 2014, Computational Statistics
  \& Data Analysis, 77, 54

\bibitem[{Perruchot {et~al.}(2011)Perruchot, Bouchy, Chazelas, D{\'\i}az,
  H{\'e}brard, Arnaud, Arnold, Avila, Delfosse, Boisse, Moreaux, Pepe, Richaud,
  Santerne, Sottile, \& T{\'e}zier}]{perruchot2011}
Perruchot, S., Bouchy, F., Chazelas, B., {et~al.} 2011, in Techniques and
  Instrumentation for Detection of Exoplanets V, Vol. 8151 (SPIE), 815115

\bibitem[{{Perruchot} {et~al.}(2008){Perruchot}, {Kohler}, {Bouchy}, {Richaud},
  {Richaud}, {Moreaux}, {Merzougui}, {Sottile}, {Hill}, {Knispel}, {Regal},
  {Meunier}, {Ilovaisky}, {Le Coroller}, {Gillet}, {Schmitt}, {Pepe}, {Fleury},
  {Sosnowska}, {Vors}, {M{\'e}gevand}, {Blanc}, {Carol}, {Point}, {Laloge}, \&
  {Brunel}}]{perruchot2008}
{Perruchot}, S., {Kohler}, D., {Bouchy}, F., {et~al.} 2008, in SPIE Conference
  Series, Vol. 7014, SPIE Conference Series

\bibitem[{{Perryman} {et~al.}(1997){Perryman}, {Lindegren}, {Kovalevsky},
  {Hoeg}, {Bastian}, {Bernacca}, {Cr{\'e}z{\'e}}, {Donati}, {Grenon}, {van
  Leeuwen}, {van der Marel}, {Mignard}, {Murray}, {Le Poole}, {Schrijver},
  {Turon}, {Arenou}, {Froeschl{\'e}}, \& {Petersen}}]{perryman1997}
{Perryman}, M.~A.~C., {Lindegren}, L., {Kovalevsky}, J., {et~al.} 1997, \aap,
  323, L49

\bibitem[{{Queloz} {et~al.}(2001){Queloz}, {Henry}, {Sivan}, {Baliunas},
  {Beuzit}, {Donahue}, {Mayor}, {Naef}, {Perrier}, \& {Udry}}]{queloz2001}
{Queloz}, D., {Henry}, G.~W., {Sivan}, J.~P., {et~al.} 2001, \aap, 379, 279

\bibitem[{{Rajpaul} {et~al.}(2015){Rajpaul}, {Aigrain}, {Osborne}, {Reece}, \&
  {Roberts}}]{rajpaul2015}
{Rajpaul}, V., {Aigrain}, S., {Osborne}, M.~A., {Reece}, S., \& {Roberts}, S.
  2015, \mnras, 452, 2269

\bibitem[{Rasmussen \& Williams(2005)}]{rasmussenwilliams2005}
Rasmussen, C.~E. \& Williams, C. K.~I. 2005, Gaussian Processes for Machine
  Learning (Adaptive Computation and Machine Learning) (The MIT Press)

\bibitem[{{Ribas} {et~al.}(2018){Ribas}, {Tuomi}, {Reiners}, {Butler},
  {Morales}, {Perger}, {Dreizler}, {Rodr{\'{\i}}guez-L{\'o}pez}, {Gonz{\'a}lez
  Hern{\'a}ndez}, {Rosich}, {Feng}, {Trifonov}, {Vogt}, {Caballero}, {Hatzes},
  {Herrero}, {Jeffers}, {Lafarga}, {Murgas}, {Nelson}, {Rodr{\'{\i}}guez},
  {Strachan}, {Tal-Or}, {Teske}, {Toledo-Padr{\'o}n}, {Zechmeister},
  {Quirrenbach}, {Amado}, {Azzaro}, {B{\'e}jar}, {Barnes}, {Berdi{\~n}as},
  {Burt}, {Coleman}, {Cort{\'e}s-Contreras}, {Crane}, {Engle}, {Guinan},
  {Haswell}, {Henning}, {Holden}, {Jenkins}, {Jones}, {Kaminski}, {Kiraga},
  {K{\"u}rster}, {Lee}, {L{\'o}pez-Gonz{\'a}lez}, {Montes}, {Morin}, {Ofir},
  {Pall{\'e}}, {Rebolo}, {Reffert}, {Schweitzer}, {Seifert}, {Shectman},
  {Staab}, {Street}, {Su{\'a}rez Mascare{\~n}o}, {Tsapras}, {Wang}, \&
  {Anglada-Escud{\'e}}}]{ribas2018}
{Ribas}, I., {Tuomi}, M., {Reiners}, A., {et~al.} 2018, \nat, 563, 365

\bibitem[{{Ricker} {et~al.}(2014){Ricker}, {Winn}, {Vanderspek}, {Latham},
  {Bakos}, {Bean}, {Berta-Thompson}, {Brown}, {Buchhave}, {Butler}, {Butler},
  {Chaplin}, {Charbonneau}, {Christensen-Dalsgaard}, {Clampin}, {Deming},
  {Doty}, {De Lee}, {Dressing}, {Dunham}, {Endl}, {Fressin}, {Ge}, {Henning},
  {Holman}, {Howard}, {Ida}, {Jenkins}, {Jernigan}, {Johnson}, {Kaltenegger},
  {Kawai}, {Kjeldsen}, {Laughlin}, {Levine}, {Lin}, {Lissauer}, {MacQueen},
  {Marcy}, {McCullough}, {Morton}, {Narita}, {Paegert}, {Palle}, {Pepe},
  {Pepper}, {Quirrenbach}, {Rinehart}, {Sasselov}, {Sato}, {Seager},
  {Sozzetti}, {Stassun}, {Sullivan}, {Szentgyorgyi}, {Torres}, {Udry}, \&
  {Villasenor}}]{ricker2014}
{Ricker}, G.~R., {Winn}, J.~N., {Vanderspek}, R., {et~al.} 2014, in Proc. SPIE,
  Vol. 9143, Space Telescopes and Instrumentation 2014: Optical, Infrared, and
  Millimeter Wave, 914320

\bibitem[{{Rivera} {et~al.}(2010){Rivera}, {Laughlin}, {Butler}, {Vogt},
  {Haghighipour}, \& {Meschiari}}]{rivera2010b}
{Rivera}, E.~J., {Laughlin}, G., {Butler}, R.~P., {et~al.} 2010, \apj, 719, 890

\bibitem[{{Roberts} {et~al.}(1987){Roberts}, {Lehar}, \&
  {Dreher}}]{roberts1987}
{Roberts}, D.~H., {Lehar}, J., \& {Dreher}, J.~W. 1987, \aj, 93, 968

\bibitem[{{Robertson} {et~al.}(2014){Robertson}, {Mahadevan}, {Endl}, \&
  {Roy}}]{robertson2014}
{Robertson}, P., {Mahadevan}, S., {Endl}, M., \& {Roy}, A. 2014, Science, 345,
  440

\bibitem[{{Rosen} {et~al.}(2016){Rosen}, {Webb}, {Watson}, {Ballet}, {Barret},
  {Braito}, {Carrera}, {Ceballos}, {Coriat}, {Della Ceca}, {Denkinson},
  {Esquej}, {Farrell}, {Freyberg}, {Gris{\'e}}, {Guillout}, {Heil},
  {Koliopanos}, {Law-Green}, {Lamer}, {Lin}, {Martino}, {Michel}, {Motch},
  {Nebot Gomez-Moran}, {Page}, {Page}, {Page}, {Pakull}, {Pye}, {Read},
  {Rodriguez}, {Sakano}, {Saxton}, {Schwope}, {Scott}, {Sturm}, {Traulsen},
  {Yershov}, \& {Zolotukhin}}]{rosen2016}
{Rosen}, S.~R., {Webb}, N.~A., {Watson}, M.~G., {et~al.} 2016, \aap, 590, A1

\bibitem[{{Sanchis-Ojeda} {et~al.}(2014){Sanchis-Ojeda}, {Rappaport}, {Winn},
  {Kotson}, {Levine}, \& {El Mellah}}]{sanchisojeda2014}
{Sanchis-Ojeda}, R., {Rappaport}, S., {Winn}, J.~N., {et~al.} 2014, \apj, 787,
  47

\bibitem[{{Schneider} {et~al.}(2011){Schneider}, {Dedieu}, {Le Sidaner},
  {Savalle}, \& {Zolotukhin}}]{schneider2011}
{Schneider}, J., {Dedieu}, C., {Le Sidaner}, P., {Savalle}, R., \&
  {Zolotukhin}, I. 2011, \aap, 532, A79+

\bibitem[{{Schwarz} {et~al.}(2016){Schwarz}, {Ginski}, {de Kok}, {Snellen},
  {Brogi}, \& {Birkby}}]{schwarz2016}
{Schwarz}, H., {Ginski}, C., {de Kok}, R.~J., {et~al.} 2016, \aap, 593, A74

\bibitem[{Sellke {et~al.}(2001)Sellke, Bayarri, \& Berger}]{sellke2001}
Sellke, T., Bayarri, M., \& Berger, J.~O. 2001, The American Statistician, 55,
  62

\bibitem[{{Snellen} {et~al.}(2015){Snellen}, {de Kok}, {Birkby}, {Brandl},
  {Brogi}, {Keller}, {Kenworthy}, {Schwarz}, \& {Stuik}}]{snellen2015}
{Snellen}, I., {de Kok}, R., {Birkby}, J.~L., {et~al.} 2015, \aap, 576, A59

\bibitem[{{Snellen} {et~al.}(2008){Snellen}, {Albrecht}, {de Mooij}, \& {Le
  Poole}}]{snellen2008}
{Snellen}, I.~A.~G., {Albrecht}, S., {de Mooij}, E.~J.~W., \& {Le Poole}, R.~S.
  2008, \aap, 487, 357

\bibitem[{{Snellen} {et~al.}(2014){Snellen}, {Brandl}, {de Kok}, {Brogi},
  {Birkby}, \& {Schwarz}}]{snellen2014}
{Snellen}, I.~A.~G., {Brandl}, B.~R., {de Kok}, R.~J., {et~al.} 2014, \nat,
  509, 63

\bibitem[{{Sparks} \& {Ford}(2002)}]{sparksford2002}
{Sparks}, W.~B. \& {Ford}, H.~C. 2002, \apj, 578, 543

\bibitem[{{Tal-Or} {et~al.}(2019){Tal-Or}, {Trifonov}, {Zucker}, {Mazeh}, \&
  {Zechmeister}}]{tal-or2019}
{Tal-Or}, L., {Trifonov}, T., {Zucker}, S., {Mazeh}, T., \& {Zechmeister}, M.
  2019, \mnras, 484, L8

\bibitem[{{van Altena} {et~al.}(1995){van Altena}, {Lee}, \&
  {Hoffleit}}]{vanaltena1995}
{van Altena}, W.~F., {Lee}, J.~T., \& {Hoffleit}, E.~D. 1995, {The general
  catalogue of trigonometric [stellar] parallaxes}

\bibitem[{{van de Kamp} \& {Lippincott}(1951)}]{vandeKamp1951}
{van de Kamp}, P. \& {Lippincott}, S.~L. 1951, \aj, 56, 49

\bibitem[{{Vaughan} {et~al.}(1978){Vaughan}, {Preston}, \& {Wilson}}]{mw}
{Vaughan}, A.~H., {Preston}, G.~W., \& {Wilson}, O.~C. 1978, \pasp, 90, 267

\bibitem[{{Wittenmyer} {et~al.}(2014){Wittenmyer}, {Tuomi}, {Butler}, {Jones},
  {Anglada-Escud{\'e}}, {Horner}, {Tinney}, {Marshall}, {Carter}, {Bailey},
  {Salter}, {O'Toole}, {Wright}, {Crane}, {Schectman}, {Arriagada}, {Thompson},
  {Minniti}, {Jenkins}, \& {Diaz}}]{wittenmyer2014}
{Wittenmyer}, R.~A., {Tuomi}, M., {Butler}, R.~P., {et~al.} 2014, \apj, 791,
  114

\bibitem[{{Wyttenbach} {et~al.}(2017){Wyttenbach}, {Lovis}, {Ehrenreich},
  {Bourrier}, {Pino}, {Allart}, {Astudillo-Defru}, {Cegla}, {Heng}, {Lavie},
  {Melo}, {Murgas}, {Santerne}, {S{\'e}gransan}, {Udry}, \&
  {Pepe}}]{wyttenbach2017}
{Wyttenbach}, A., {Lovis}, C., {Ehrenreich}, D., {et~al.} 2017, \aap, 602, A36

\bibitem[{{Zechmeister} \& {K{\"u}rster}(2009)}]{zechmeisterkurster2009}
{Zechmeister}, M. \& {K{\"u}rster}, M. 2009, \aap, 496, 577

\end{thebibliography}

\begin{appendix}
\section{Additional plots and tables\label{appendix.freqperiodograms}}

\begin{figure}
\input{figures/Gl411_GLSfreq.pgf}
\caption{Same as Fig.~\ref{fig.periodograms} but using a linear scale in frequency in the $x$-axis.}
\label{fig.periodogramsfreq}
\end{figure}

\begin{figure*}
\input{figures/Gl411_GLSfreqancillary.pgf}
\caption{Same as Fig.~\ref{fig.GLSanci} but using a linear scale in frequency in the $x$-axis.}
\label{fig.GLSancifreq}
\end{figure*}

\begin{table*}[t]
{\centering
\caption{Parameter prior distributions. \label{table.priors}}            
\begin{tabular}{l l c}        
\hline\hline                 
\noalign{\smallskip}

	&&Prior\\
\multicolumn{2}{c}{Stellar parameters} 		\\
\hline
\noalign{\smallskip}
Systemic radial velocity, $\gamma_0$\tablefootmark{$^\dagger$} & [\kms] &$U(-84.79, -84.59)$\\
Parallax, $\pi$\tablefootmark{$^\dagger$} & [mas] &$N(392.64, 0.67)$\\
Proper motion in R.A., $\mu_\alpha$\tablefootmark{$^\dagger$} & [mas/yr] &$N(-580.27, 0.62)$\\
Proper motion in Dec., $\mu_\delta$\tablefootmark{$^\dagger$} & [mas/yr] &$N(-4765.85, 0.64)$\\

\noalign{\smallskip}
\multicolumn{2}{c}{ Orbital parameters } \\
\hline
\noalign{\smallskip}
Orbital period, $P$ 	&[days]	 		&$J(1, 600)$\\

RV amplitude, $K$ 	&[\ms]			&$mJ(1, 10)$	\\

$\sqrt{e}\cos{\omega}$ &$--$			&$U(-1, 1)$\\

$\sqrt{e}\sin{\omega}$ &$--$			&$U(-1, 1)$\\

$\left(\sqrt{e}\sin{\omega}\right)^2 + \left(\sqrt{e}\cos{\omega}\right)^2$& $--$ 			&$U(0, 1)$\\

Mean longitude at epoch, $\lambda_0$ &[deg] 	&$U(-180, 180)$\\

\noalign{\smallskip}
\multicolumn{2}{c}{Drift parameters } \\
\hline
\noalign{\smallskip}
Constant acceleration, $\gamma_1$				&[\ms/yr]	& $U(-10, 10)$\\

Linear acceleration, $\gamma_2$				&[m$^2$s$^{-2}$/yr$^2$]	& $U(-10, 10)$\\

\noalign{\smallskip}
\multicolumn{2}{c}{Noise model} 		\\
\hline
\noalign{\smallskip}
Covariance amplitude, $A$ & [m s$^{-1}$] & $mJ(1, 10)$\\

Log-characteristic timesale, $\log_{10}\tau$ & [d] & $U(-1, 5)$\\

Extra white-noise amplitude, $\sigma_J$ & [m s$^{-1}$] &$mJ(1, 10)$\\

Zero-point proportionaly constant, $a_c$ &$--$ & $U(-10, 10)$\\

\noalign{\smallskip}
\hline

\end{tabular}

\tablefoot{\\
\tablefootmark{$^\dagger$} at the epoch of Hipparcos data, J1991.25, $JD = 2\,448\,349.0625$\\
$U(x_\text{min};  x_\text{max})$: uniform distribution between $x_\text{min}$ and $x_\text{max}$.\\
$J(x_\text{min};  x_\text{max})$: Jeffreys (log-flat) distribution between $x_\text{min}$ and $x_\text{max}$.\\
$mJ(x_{0};  x_\text{max})$: Modified Jeffreys (log-flat) distribution, between 0 and $x_\text{max}$, $p(x)\mathrm{d}x = \frac{\mathrm{d}x}{x_0\left(1 + x/x_0\right)}\frac{1}{\log\left(1 + x_\text{max}/x_0\right)}$\\ 
$N(\mu; \sigma)$: normal distribution with mean $\mu$ and standard deviation $\sigma$.\\
}

}
\end{table*}

\section{Details on the ground-based APT photometry. \label{sect.apt}}

\begin{table*}[t]
\caption{Results of the Photometric Analysis of Gl411}
\begin{tabular}{ccccccccc}
\hline\hline
& Photometric & HJD Range & & $<(V-C)>$ & Period & Amplitude & $<(K-C)>$ & $\sigma_{K-C}$\\
Season & Band & (HJD $-$ 2,400,000) & $N_\mathrm{obs}$ & (mag) & (days) & (mag) & (mag) & (mag) \\
\hline
2011 & $V$ & 55,652--55,728 & 156 & 0.0338 & 60.16 $\pm$ 1.01 & 0.0024  & $-$1.7177 & 0.0050 \\
2011 & $B$ & 55,652--55,728 & 176 & 1.1060 & 65.62 $\pm$ 4.48 & 0.0019  & $-$1.0864 & 0.0054 \\
2012 & $V$ & 55,849--56,087 & 390 & 0.0321 & 59.42 $\pm$ 0.67 & 0.0075  & $-$1.7160 & 0.0043 \\
2012 & $B$ & 55,851--56,087 & 386 & 1.1052 & 60.90 $\pm$ 0.74 & 0.0064  & $-$1.0846 & 0.0037 \\
2013 & $V$ & 56,224--56,397 &  55 & 0.0368 & 50.81 $\pm$ 0.93 & 0.0076  & $-$1.7125 & 0.0051 \\
2013 & $B$ & 56,227--56,397 &  49 & 1.1082 & 48.95 $\pm$ 1.14 & 0.0084  & $-$1.0814 & 0.0050 \\
2014 & $V$ & 56,670--56,821 &  60 & 0.0375 & 61.31 $\pm$ 1.39 & 0.0052  & $-$1.7139 & 0.0032 \\
2014 & $B$ & 56,670--56,821 &  63 & 1.1068 & 62.42 $\pm$ 1.71 & 0.0063  & $-$1.0835 & 0.0030 \\
2015 & $V$ & 56,969--57,187 & 128 & 0.0337 &      \nodata     & \nodata & $-$1.7143 & 0.0039 \\
2015 & $B$ & 56,968--57,187 & 123 & 1.1026 & 56.09 $\pm$ 0.72 & 0.0029  & $-$1.0842 & 0.0027 \\
2016 & $V$ & 57,329--57,556 &  97 & 0.0343 & 51.26 $\pm$ 0.76 & 0.0037  & $-$1.7145 & 0.0033 \\
2016 & $B$ & 57,329--57,556 & 114 & 1.1025 & 53.67 $\pm$ 0.75 & 0.0039  & $-$1.0857 & 0.0036 \\
2017 & $V$ & 57,698--57,922 & 119 & 0.0323 &      \nodata     & \nodata & $-$1.7142 & 0.0030 \\
2017 & $B$ & 57,698--57,922 & 117 & 1.1005 & 55.64 $\pm$ 1.08 & 0.0032  & $-$1.0851 & 0.0038 \\
2018 & $V$ & 58,071--58,277 &  50 & 0.0369 &      \nodata     & \nodata & $-$1.7151 & 0.0044 \\
2018 & $B$ & 58,071--58,275 &  55 & 1.1052 &      \nodata     & \nodata & $-$1.0842 & 0.0041 \\
\hline
\end{tabular}
\label{table.photometry}
\end{table*}

The 0.40~m automatic photoelectric telescope (APT) at Fairborn Observatory in southern Arizona uses a temperature-stabilized EMI 9924B photomultiplier tube to 
measure stellar count rates successively through Johnson $B$ and $V$ filters. 
Gl411 was measured each clear night in the following sequence, termed a 
group observation: {\it K,S,C,V,C,V,C,V,C,S,K}, in which $K$ is the check 
star HD~94497 ($V=5.73$, $B-V=1.03$, G7III:), $C$ is the comparison star 
HD~95485 ($V=7.45$, $B-V=0.40$, F0V), $V$ is Gl411 ($V=7.45$, $B-V=0.40$, 
F0V), and $S$ is a sky reading.  Three $V-C$ and two $K-C$ differential 
magnitudes are formed from each sequence and averaged together to create group 
means.  Group mean differential magnitudes with internal standard deviations 
greater than 0.01~mag were rejected to filter observations taken under 
non-photometric conditions.  The surviving group means were corrected for 
differential extinction and transformed to the Johnson system.  The external 
precision of these group means, based on standard deviations for pairs of 
constant stars, is typically $\sim$0.003--0.005~mag on good nights.

Further information on the observations are given in Table~\ref{table.photometry}, which presents the date range of observations for each season as well as the number of measurements obtained in columns 3 and 4, respectively.

Column~5 lists the mean differential magnitudes for each observing season.  
The $V$ and $B$ seasonal means scatter about their grand means with standard 
deviations of 0.0021 and 0.0026 mag, respectively.  This suggests small 
brightness variations from year to year may be occurring, but our data set is 
not long enough to comment on possible brightness (magnetic) cycles.

Finally, the seasonal mean differential magnitudes of the check minus 
comparison star ($K-C$) and the seasonal standard deviations are given in
columns 8 and 9, respectively.  The seasonal standard deviations in column~9
are consistent with our measurement precision, indicating both the comparison 
and check stars are constant from night to night.  The standard deviations of
the $V$ and $B$ seasonal mean magnitudes in column~8 scatter about their
grand mean with a standard deviation of only 0.0015 mag in both pass bands,
indicating the comparison and checks stars are also constant over longer time 
scales. Further information on the operation of our APTs and processing of the data can be 
found in \citet{henry1995a}, \citet{henry1995b}, \citet{henry1999}, and \citet{eaton2003}.

\section{Posterior summaries for individual models.}
In this section we present tables with posterior summaries for all individual models tested. For each model, the maximum-a-posteriori estimate is reported. The number after the $\pm$ symbol is the standard deviation for each model. Additionally, the 95\%-highest density interval computed on the marginal samples from each parameter is reported in a separate line.
\begin{sidewaystable*}
\caption{Posterior summaries for individual models without a Keplerian component and assuming white noise.}
\label{table.posteriorsk0wn}
\begin{tabular}{llccc}
\hline
\hline\noalign{\smallskip}
&& \multicolumn{3}{c}{\bf Models}\\ 
&& k0d0wn & k0d1wn & k0d2wn\\ 
\hline\noalign{\smallskip}
\multicolumn{3}{l}{\bf Stellar parameters}\\ 
\hline\noalign{\smallskip}
Systemic velocity\tablefootmark{$\dagger$}, $\gamma_0$ & [km s$^{-1}$] & $-84.64538 \pm 0.00021$ & $-84.64521 \pm 0.00021$ & $-84.64587 \pm 0.00036$  \\ 
&& $[-84.64579, -84.64494]$& $[-84.64564, -84.64482]$& $[-84.64654, -84.64505]$ \\ 
\noalign{\smallskip}
Parallax\tablefootmark{$\dagger$}, $\pi$ & [mas] & $392.74 \pm 0.67$ & $392.69 \pm 0.67$ & $392.59 \pm 0.67$  \\ 
&& $[391.33, 394.01]$& $[391.30, 394.05]$& $[391.34, 393.98]$ \\ 
\noalign{\smallskip}
Proper motion (RA)\tablefootmark{$\dagger$}, $\mu_\alpha$ & [mas yr$^{-1}$] & $-580.26 \pm 0.62$ & $-580.30 \pm 0.62$ & $-580.28 \pm 0.62$  \\ 
&& $[-581.51, -579.07]$& $[-581.57, -579.00]$& $[-581.60, -579.06]$ \\ 
\noalign{\smallskip}
Proper motion (DEC)\tablefootmark{$\dagger$}, $\mu_\delta$ & [mas yr$^{-1}$] & $-4765.83 \pm 0.64$ & $-4765.79 \pm 0.64$ & $-4765.82 \pm 0.64$  \\ 
&& $[-4767.11, -4764.50]$& $[-4767.09, -4764.57]$& $[-4767.14, -4764.51]$ \\ 
\noalign{\smallskip}
\hline\noalign{\smallskip}
\multicolumn{3}{l}{\bf Drift parameters}\\ 
\hline\noalign{\smallskip}
Constant acceleration, $\gamma_1$ & [m s$^{-1}$ yr$^{-1}$] & --& $-0.386 \pm 0.094$ & $-0.370 \pm 0.094$  \\ 
&& & $[-0.575, -0.196]$& $[-0.553, -0.183]$ \\ 
\noalign{\smallskip}
Linear acceleration, $\gamma_2$ & [m$^2$ s$^{-2}$ yr$^{-2}$] & --& --& $0.127 \pm 0.064$  \\ 
&& & & $[-0.007, 0.250]$ \\ 
\noalign{\smallskip}
\noalign{\smallskip}
\hline\noalign{\smallskip}
\multicolumn{3}{l}{\bf Instrumental parameters}\\ 
\hline\noalign{\smallskip}
Extra white-noise amplitude, $\sigma_J$ & [m s$^{-1}$] & $2.12 \pm 0.17$ & $1.98 \pm 0.17$ & $1.93 \pm 0.17$  \\ 
&& $[1.82, 2.50]$& $[1.66, 2.34]$& $[1.65, 2.31]$ \\ 
\noalign{\smallskip}
Zero-point proportionaly constant, $a_c$ & -- & $0.901 \pm 0.077$ & $0.892 \pm 0.073$ & $0.812 \pm 0.084$  \\ 
&& $[0.753, 1.066]$& $[0.751, 1.041]$& $[0.641, 0.977]$ \\ 
\noalign{\smallskip}
\hline
\end{tabular}
\tablefoot{\tablefootmark{$\dagger$}Measured at Hipparcos epoch}.
\end{sidewaystable*}

\begin{sidewaystable*}
\caption{Posterior summaries for individual models without a Keplerian component and assuming correlated noise.}
\label{table.posteriorsk0rn}
\begin{tabular}{llccc}
\hline
\hline\noalign{\smallskip}
&& \multicolumn{3}{c}{\bf Models}\\ 
&& k0d0rn & k0d1rn & k0d2rn\\ 
\hline\noalign{\smallskip}
\multicolumn{3}{l}{\bf Stellar parameters}\\ 
\hline\noalign{\smallskip}
Systemic velocity\tablefootmark{$\dagger$}, $\gamma_0$ & [km s$^{-1}$] & $-84.64520 \pm 0.00031$ & $-84.64535 \pm 0.00028$ & $-84.64630 \pm 0.00051$  \\ 
&& $[-84.64589, -84.64462]$& $[-84.64594, -84.64474]$& $[-84.64733, -84.64528]$ \\ 
\noalign{\smallskip}
Parallax\tablefootmark{$\dagger$}, $\pi$ & [mas] & $392.49 \pm 0.67$ & $392.58 \pm 0.67$ & $392.35 \pm 0.67$  \\ 
&& $[391.30, 394.03]$& $[391.29, 394.02]$& $[391.21, 393.95]$ \\ 
\noalign{\smallskip}
Proper motion (RA)\tablefootmark{$\dagger$}, $\mu_\alpha$ & [mas yr$^{-1}$] & $-580.24 \pm 0.62$ & $-580.13 \pm 0.62$ & $-580.31 \pm 0.62$  \\ 
&& $[-581.58, -579.07]$& $[-581.58, -579.06]$& $[-581.46, -578.97]$ \\ 
\noalign{\smallskip}
Proper motion (DEC)\tablefootmark{$\dagger$}, $\mu_\delta$ & [mas yr$^{-1}$] & $-4765.79 \pm 0.64$ & $-4765.94 \pm 0.64$ & $-4765.99 \pm 0.64$  \\ 
&& $[-4767.12, -4764.59]$& $[-4767.10, -4764.60]$& $[-4767.17, -4764.58]$ \\ 
\noalign{\smallskip}
\hline\noalign{\smallskip}
\multicolumn{3}{l}{\bf Drift parameters}\\ 
\hline\noalign{\smallskip}
Constant acceleration, $\gamma_1$ & [m s$^{-1}$ yr$^{-1}$] & --& $-0.49 \pm 0.13$ & $-0.42 \pm 0.13$  \\ 
&& & $[-0.75, -0.21]$& $[-0.69, -0.15]$ \\ 
\noalign{\smallskip}
Linear acceleration, $\gamma_2$ & [m$^2$ s$^{-2}$ yr$^{-2}$] & --& --& $0.207 \pm 0.091$  \\ 
&& & & $[0.007, 0.380]$ \\ 
\noalign{\smallskip}
\noalign{\smallskip}
\hline\noalign{\smallskip}
\multicolumn{3}{l}{\bf Instrumental parameters}\\ 
\hline\noalign{\smallskip}
Covariance amplitude, $A$ & [m s$^{-1}$] & $2.07 \pm 0.29$ & $1.87 \pm 0.26$ & $1.84 \pm 0.26$  \\ 
&& $[1.47, 2.69]$& $[1.43, 2.51]$& $[1.41, 2.43]$ \\ 
\noalign{\smallskip}
Log-characteristic timesale, $\log_{10}\tau$ & [d] & $1.36 \pm 0.44$ & $1.33 \pm 0.33$ & $1.36 \pm 0.33$  \\ 
&& $[0.82, 2.61]$& $[0.79, 2.22]$& $[0.84, 2.19]$ \\ 
\noalign{\smallskip}
Extra white-noise amplitude, $\sigma_J$ & [m s$^{-1}$] & $0.77 \pm 0.39$ & $0.70 \pm 0.36$ & $0.69 \pm 0.36$  \\ 
&& $[0.00, 1.49]$& $[0.00, 1.36]$& $[0.00, 1.29]$ \\ 
\noalign{\smallskip}
Zero-point proportionaly constant, $a_c$ & -- & $0.89 \pm 0.10$ & $0.833 \pm 0.091$ & $0.72 \pm 0.11$  \\ 
&& $[0.64, 1.05]$& $[0.633, 1.011]$& $[0.47, 0.92]$ \\ 
\noalign{\smallskip}
\hline
\end{tabular}
\tablefoot{\tablefootmark{$\dagger$}Measured at Hipparcos epoch}.
\end{sidewaystable*}

\begin{sidewaystable*}
\caption{Posterior summaries for individual models with a Keplerian component and assuming white noise.}
\label{table.posteriorsk1wn}
\begin{tabular}{llccc}
\hline
\hline\noalign{\smallskip}
&& \multicolumn{3}{c}{\bf Models}\\ 
&& k1d0wn & k1d1wn & k1d2wn\\ 
\hline\noalign{\smallskip}
\multicolumn{3}{l}{\bf Stellar parameters}\\ 
\hline\noalign{\smallskip}
Systemic velocity\tablefootmark{$\dagger$}, $\gamma_0$ & [km s$^{-1}$] & $-84.64539 \pm 0.00019$ & $-84.64527 \pm 0.00019$ & $-84.64594 \pm 0.00032$  \\ 
&& $[-84.64579, -84.64501]$& $[-84.64563, -84.64488]$& $[-84.64648, -84.64522]$ \\ 
\noalign{\smallskip}
Parallax\tablefootmark{$\dagger$}, $\pi$ & [mas] & $392.83 \pm 0.67$ & $392.63 \pm 0.67$ & $392.48 \pm 0.67$  \\ 
&& $[391.34, 394.03]$& $[391.35, 394.02]$& $[391.31, 394.05]$ \\ 
\noalign{\smallskip}
Proper motion (RA)\tablefootmark{$\dagger$}, $\mu_\alpha$ & [mas yr$^{-1}$] & $-580.23 \pm 0.62$ & $-580.24 \pm 0.62$ & $-580.39 \pm 0.62$  \\ 
&& $[-581.48, -579.01]$& $[-581.47, -579.04]$& $[-581.54, -579.10]$ \\ 
\noalign{\smallskip}
Proper motion (DEC)\tablefootmark{$\dagger$}, $\mu_\delta$ & [mas yr$^{-1}$] & $-4765.64 \pm 0.64$ & $-4765.91 \pm 0.64$ & $-4765.88 \pm 0.64$  \\ 
&& $[-4767.20, -4764.61]$& $[-4767.14, -4764.61]$& $[-4767.12, -4764.56]$ \\ 
\noalign{\smallskip}
\hline\noalign{\smallskip}
\multicolumn{3}{l}{\bf Planet parameters}\\ 
\hline\noalign{\smallskip}
Orbital period, $P$ & [d] & $12.9439 \pm 0.0056$ & $12.9444 \pm 0.0081$ & $12.9448 \pm 0.0076$  \\ 
&& $[12.9292, 12.9520]$& $[12.9332, 12.9673]$& $[12.9341, 12.9656]$ \\ 
\noalign{\smallskip}
RV amplitude, $K$ & [m s$^{-1}$] & $1.88 \pm 0.32$ & $1.90 \pm 0.35$ & $1.91 \pm 0.31$  \\ 
&& $[1.25, 2.58]$& $[1.01, 2.40]$& $[1.09, 2.39]$ \\ 
\noalign{\smallskip}
$\sqrt{e} \sin{\omega}$ &  & $-0.59 \pm 0.18$ & $-0.66 \pm 0.23$ & $-0.65 \pm 0.23$  \\ 
&& $[-0.83, -0.14]$& $[-0.83, 0.04]$& $[-0.82, 0.02]$ \\ 
\noalign{\smallskip}
$\sqrt{e} \cos{\omega}$ &  & $-0.24 \pm 0.27$ & $-0.13 \pm 0.28$ & $-0.11 \pm 0.27$  \\ 
&& $[-0.66, 0.38]$& $[-0.60, 0.46]$& $[-0.63, 0.41]$ \\ 
\noalign{\smallskip}
Mean longitude at epoch, $\lambda_0$ & [deg] & $-52.9 \pm 10.1$ & $-46.5 \pm 11.0$ & $-45.3 \pm 10.5$  \\ 
&& $[-73.6, -32.5]$& $[-69.4, -24.6]$& $[-66.9, -24.6]$ \\ 
\noalign{\smallskip}
\noalign{\smallskip}
\hline\noalign{\smallskip}
\multicolumn{3}{l}{\bf Drift parameters}\\ 
\hline\noalign{\smallskip}
Constant acceleration, $\gamma_1$ & [m s$^{-1}$ yr$^{-1}$] & --& $-0.340 \pm 0.089$ & $-0.246 \pm 0.088$  \\ 
&& & $[-0.498, -0.149]$& $[-0.489, -0.141]$ \\ 
\noalign{\smallskip}
Linear acceleration, $\gamma_2$ & [m$^2$ s$^{-2}$ yr$^{-2}$] & --& --& $0.144 \pm 0.056$  \\ 
&& & & $[0.025, 0.247]$ \\ 
\noalign{\smallskip}
\noalign{\smallskip}
\hline\noalign{\smallskip}
\multicolumn{3}{l}{\bf Instrumental parameters}\\ 
\hline\noalign{\smallskip}
Extra white-noise amplitude, $\sigma_J$ & [m s$^{-1}$] & $1.69 \pm 0.17$ & $1.54 \pm 0.16$ & $1.45 \pm 0.16$  \\ 
&& $[1.42, 2.10]$& $[1.31, 1.98]$& $[1.28, 1.94]$ \\ 
\noalign{\smallskip}
Zero-point proportionaly constant, $a_c$ & -- & $0.885 \pm 0.068$ & $0.907 \pm 0.066$ & $0.791 \pm 0.074$  \\ 
&& $[0.762, 1.034]$& $[0.771, 1.034]$& $[0.667, 0.960]$ \\ 
\noalign{\smallskip}
\hline
\end{tabular}
\tablefoot{\tablefootmark{$\dagger$}Measured at Hipparcos epoch}.
\end{sidewaystable*}

\begin{sidewaystable*}
\caption{Posterior summaries for individual models without a Keplerian component and assuming correlated noise.}
\label{table.posteriorsk1rn}
\begin{tabular}{llccc}
\hline
\hline\noalign{\smallskip}
&& \multicolumn{3}{c}{\bf Models}\\ 
&& k1d0rn & k1d1rn & k1d2rn\\ 
\hline\noalign{\smallskip}
\multicolumn{3}{l}{\bf Stellar parameters}\\ 
\hline\noalign{\smallskip}
Systemic velocity\tablefootmark{$\dagger$}, $\gamma_0$ & [km s$^{-1}$] & $-84.64523 \pm 0.00036$ & $-84.64542 \pm 0.00033$ & $-84.64642 \pm 0.00056$  \\ 
&& $[-84.64596, -84.64452]$& $[-84.64610, -84.64472]$& $[-84.64767, -84.64537]$ \\ 
\noalign{\smallskip}
Parallax\tablefootmark{$\dagger$}, $\pi$ & [mas] & $392.46 \pm 0.67$ & $392.84 \pm 0.67$ & $392.59 \pm 0.67$  \\ 
&& $[391.35, 394.10]$& $[391.30, 394.04]$& $[391.22, 393.98]$ \\ 
\noalign{\smallskip}
Proper motion (RA)\tablefootmark{$\dagger$}, $\mu_\alpha$ & [mas yr$^{-1}$] & $-580.30 \pm 0.62$ & $-580.21 \pm 0.62$ & $-580.07 \pm 0.62$  \\ 
&& $[-581.52, -578.97]$& $[-581.58, -579.06]$& $[-581.58, -579.03]$ \\ 
\noalign{\smallskip}
Proper motion (DEC)\tablefootmark{$\dagger$}, $\mu_\delta$ & [mas yr$^{-1}$] & $-4765.83 \pm 0.64$ & $-4765.84 \pm 0.64$ & $-4766.22 \pm 0.64$  \\ 
&& $[-4767.05, -4764.52]$& $[-4767.07, -4764.50]$& $[-4767.07, -4764.50]$ \\ 
\noalign{\smallskip}
\hline\noalign{\smallskip}
\multicolumn{3}{l}{\bf Planet parameters}\\ 
\hline\noalign{\smallskip}
Orbital period, $P$ & [d] & $12.9465 \pm 0.0071$ & $12.9476 \pm 0.0080$ & $12.9477 \pm 0.0074$  \\ 
&& $[12.9374, 12.9662]$& $[12.9401, 12.9715]$& $[12.9399, 12.9688]$ \\ 
\noalign{\smallskip}
RV amplitude, $K$ & [m s$^{-1}$] & $1.75 \pm 0.24$ & $1.76 \pm 0.23$ & $1.70 \pm 0.23$  \\ 
&& $[1.19, 2.14]$& $[1.11, 2.03]$& $[1.14, 2.08]$ \\ 
\noalign{\smallskip}
$\sqrt{e} \sin{\omega}$ &  & $-0.58 \pm 0.22$ & $-0.57 \pm 0.24$ & $-0.51 \pm 0.23$  \\ 
&& $[-0.72, 0.12]$& $[-0.67, 0.22]$& $[-0.67, 0.22]$ \\ 
\noalign{\smallskip}
$\sqrt{e} \cos{\omega}$ &  & $-0.09 \pm 0.23$ & $-0.13 \pm 0.24$ & $-0.19 \pm 0.23$  \\ 
&& $[-0.56, 0.35]$& $[-0.51, 0.39]$& $[-0.54, 0.35]$ \\ 
\noalign{\smallskip}
Mean longitude at epoch, $\lambda_0$ & [deg] & $-42.1 \pm 9.0$ & $-41.8 \pm 9.5$ & $-43.4 \pm 9.1$  \\ 
&& $[-60.3, -22.1]$& $[-56.4, -18.8]$& $[-58.9, -22.6]$ \\ 
\noalign{\smallskip}
\noalign{\smallskip}
\hline\noalign{\smallskip}
\multicolumn{3}{l}{\bf Drift parameters}\\ 
\hline\noalign{\smallskip}
Constant acceleration, $\gamma_1$ & [m s$^{-1}$ yr$^{-1}$] & --& $-0.49 \pm 0.15$ & $-0.37 \pm 0.15$  \\ 
&& & $[-0.85, -0.22]$& $[-0.71, -0.08]$ \\ 
\noalign{\smallskip}
Linear acceleration, $\gamma_2$ & [m$^2$ s$^{-2}$ yr$^{-2}$] & --& --& $0.217 \pm 0.098$  \\ 
&& & & $[0.043, 0.438]$ \\ 
\noalign{\smallskip}
\noalign{\smallskip}
\hline\noalign{\smallskip}
\multicolumn{3}{l}{\bf Instrumental parameters}\\ 
\hline\noalign{\smallskip}
Covariance amplitude, $A$ & [m s$^{-1}$] & $1.94 \pm 0.32$ & $1.76 \pm 0.28$ & $1.56 \pm 0.27$  \\ 
&& $[1.38, 2.73]$& $[1.30, 2.42]$& $[1.24, 2.32]$ \\ 
\noalign{\smallskip}
Log-characteristic timesale, $\log_{10}\tau$ & [d] & $2.20 \pm 0.25$ & $2.17 \pm 0.24$ & $2.11 \pm 0.25$  \\ 
&& $[1.78, 2.82]$& $[1.63, 2.69]$& $[1.68, 2.67]$ \\ 
\noalign{\smallskip}
Extra white-noise amplitude, $\sigma_J$ & [m s$^{-1}$] & $0.48 \pm 0.29$ & $0.32 \pm 0.27$ & $0.12 \pm 0.26$  \\ 
&& $[0.00, 0.99]$& $[0.00, 0.94]$& $[0.00, 0.90]$ \\ 
\noalign{\smallskip}
Zero-point proportionaly constant, $a_c$ & -- & $0.78 \pm 0.12$ & $0.79 \pm 0.11$ & $0.72 \pm 0.12$  \\ 
&& $[0.52, 0.99]$& $[0.55, 0.99]$& $[0.39, 0.86]$ \\ 
\noalign{\smallskip}
\hline
\end{tabular}
\tablefoot{\tablefootmark{$\dagger$}Measured at Hipparcos epoch}.
\end{sidewaystable*}

\end{appendix}

\end{document}